\PassOptionsToPackage{unicode}{hyperref}
\PassOptionsToPackage{hyphens}{url}
\PassOptionsToPackage{dvipsnames,svgnames,x11names}{xcolor}
\documentclass[12pt]{article}
\usepackage{amsmath,amssymb}
\usepackage{orcidlink}
\usepackage{iftex}
\ifPDFTeX
  \usepackage[T1]{fontenc}
  \usepackage[utf8]{inputenc}
  \usepackage{textcomp} 
\else 
  \usepackage{unicode-math}
  \defaultfontfeatures{Scale=MatchLowercase}
  \defaultfontfeatures[\rmfamily]{Ligatures=TeX,Scale=1}
\fi
\usepackage{lmodern}
\ifPDFTeX\else  
\fi
\IfFileExists{upquote.sty}{\usepackage{upquote}}{}
\IfFileExists{microtype.sty}{
  \usepackage[]{microtype}
  \UseMicrotypeSet[protrusion]{basicmath} 
}{}
\makeatletter
\@ifundefined{KOMAClassName}{
  \IfFileExists{parskip.sty}{%
    \usepackage{parskip}
  }{
    \setlength{\parindent}{0pt}
    \setlength{\parskip}{6pt plus 2pt minus 1pt}}
}{
  \KOMAoptions{parskip=half}}
\makeatother
\usepackage{xcolor}
\usepackage{enumitem}
\usepackage{dsfont}
\makeatletter
\ifx\paragraph\undefined\else
  \let\oldparagraph\paragraph
  \renewcommand{\paragraph}{
    \@ifstar
      \xxxParagraphStar
      \xxxParagraphNoStar
  }
  \newcommand{\xxxParagraphStar}[1]{\oldparagraph*{#1}\mbox{}}
  \newcommand{\xxxParagraphNoStar}[1]{\oldparagraph{#1}\mbox{}}
\fi
\ifx\subparagraph\undefined\else
  \let\oldsubparagraph\subparagraph
  \renewcommand{\subparagraph}{
    \@ifstar
      \xxxSubParagraphStar
      \xxxSubParagraphNoStar
  }
  \newcommand{\xxxSubParagraphStar}[1]{\oldsubparagraph*{#1}\mbox{}}
  \newcommand{\xxxSubParagraphNoStar}[1]{\oldsubparagraph{#1}\mbox{}}
\fi
\makeatother

\newtheorem{prop}{Proposition}

\newtheorem{result}{Result}

\newlist{assumption}{enumerate}{1}
\setlist[assumption]{label=\textbf{(A\arabic*)}, start=1, leftmargin=2em}

\newcommand{\mB}{\mathcal B}

\newcommand{\mU}{\mathcal U}
\newcommand{\mT}{\mathcal T}
\newcommand{\mZ}{\mathcal Z}
\newcommand{\mJ}{\mathcal J}
\newcommand{\mV}{\mathcal V}
\newcommand{\mC}{\mathcal C}
\newcommand{\Hl}{L^2(G)}

\newcommand{\normS}{\|S\|_{G}}
\newcommand{\eps}{\epsilon}

\newcommand{\intlu}{\int_{\mathcal{X}}}

\newcommand{\summ}{\sum_{i=1}^m}
\newcommand{\sumn}{\sum_{i=1}^n}
\newcommand{\Sd}{S^\dagger}
\newcommand{\bt}{\beta}
\newcommand{\btt}{\tilde\beta}

\newcommand{\bth}{\widehat{\beta}}
\newcommand{\bts}{\beta^*}
\newcommand{\del}{\delta}
\newcommand{\tht}{\theta}
\newcommand{\delh}{\widehat{\delta}}
\newcommand{\thth}{\widehat{\theta}}
\newcommand{\nut}{\tilde{\nu}}

\newcommand{\R}{\mathbb{R}}
\newcommand{\al}{\alpha}
\newcommand{\als}{\alpha^*}

\newcommand{\alh}{\widehat{\alpha}}

\newcommand{\Rp}{\mathbb{R}^p}
\newcommand{\mP}{\mathbb{P}}
\newcommand{\mE}{\mathbb{E}}
\newcommand{\mN}{\mathcal{N}}
\newcommand{\sig}{\sigma}
\newcommand{\sigh}{\widehat{\sigma}}
\newcommand{\pone}{\partial_1}
\newcommand{\ptwo}{\partial_2}
\newcommand{\pb}{\partial_\beta}
\newcommand{\ppb}{\partial^2_\beta}

\newcommand{\p}{\partial}
\newcommand{\pp}{\partial^2}
\newcommand{\Ts}{T^*}
\newcommand{\Tc}{T^{\circ}}

\newcommand{\pconv}{\stackrel{\mP}\longrightarrow}
\newcommand{\dconv}{\stackrel{d}\longrightarrow}

\usepackage{longtable,booktabs,array}
\usepackage{calc} 
\usepackage{etoolbox}
\makeatletter
\patchcmd\longtable{\par}{\if@noskipsec\mbox{}\fi\par}{}{}
\makeatother
\IfFileExists{footnotehyper.sty}{\usepackage{footnotehyper}}{\usepackage{footnote}}
\makesavenoteenv{longtable}
\usepackage{graphicx}
\makeatletter
\def\maxwidth{\ifdim\Gin@nat@width>\linewidth\linewidth\else\Gin@nat@width\fi}
\def\maxheight{\ifdim\Gin@nat@height>\textheight\textheight\else\Gin@nat@height\fi}
\makeatother
\setkeys{Gin}{width=\maxwidth,height=\maxheight,keepaspectratio}
\makeatletter
\def\fps@figure{htbp}
\makeatother

\makeatletter
\@ifpackageloaded{caption}{}{\usepackage{caption}}
\AtBeginDocument{%
\ifdefined\contentsname
  \renewcommand*\contentsname{Table of contents}
\else
  \newcommand\contentsname{Table of contents}
\fi
\ifdefined\listfigurename
  \renewcommand*\listfigurename{List of Figures}
\else
  \newcommand\listfigurename{List of Figures}
\fi
\ifdefined\listtablename
  \renewcommand*\listtablename{List of Tables}
\else
  \newcommand\listtablename{List of Tables}
\fi
\ifdefined\figurename
  \renewcommand*\figurename{Figure}
\else
  \newcommand\figurename{Figure}
\fi
\ifdefined\tablename
  \renewcommand*\tablename{Table}
\else
  \newcommand\tablename{Table}
\fi
}
\@ifpackageloaded{float}{}{\usepackage{float}}
\floatstyle{ruled}
\@ifundefined{c@chapter}{\newfloat{codelisting}{h}{lop}}{\newfloat{codelisting}{h}{lop}[chapter]}
\floatname{codelisting}{Listing}

\makeatother
\makeatletter
\@ifpackageloaded{caption}{}{\usepackage{caption}}
\@ifpackageloaded{subcaption}{}{\usepackage{subcaption}}
\makeatother

\ifLuaTeX
  \usepackage{selnolig}
\fi
\usepackage[]{natbib}
\usepackage{bookmark}

\IfFileExists{xurl.sty}{\usepackage{xurl}}{}
\hypersetup{
  pdftitle={Detecting signals under unknown background: a geometric perspective},
  pdfauthor={Author 1; Author 2},
  pdfkeywords={3 to 6 keywords, that do not appear in the title},
  colorlinks=true,
  linkcolor={blue},
  filecolor={Maroon},
  citecolor={Blue},
  urlcolor={Blue},
  pdfcreator={LaTeX via pandoc}}

\newcommand{\anon}{1}

\begin{document}

\def\spacingset#1{\renewcommand{\baselinestretch}%
{#1}\small\normalsize} \spacingset{1}


\if1\anon
{
  \title{\bf Compensator-based inference for signal detection under unknown background}
 \author{Aritra Banerjee \orcidlink{0009-0003-8777-5529} and Sara Algeri \orcidlink{0000-0001-7366-3866}\thanks{
   The authors gratefully acknowledge \textit{the Mid-Career Faculty Research Award and the Errett W. McDiarmid Graduate Fellowship provided by the College of Liberal Arts at the University of Minnesota.}}\hspace{.2cm}\\
   \emph{School of Statistics, University of Minnesota, Minneapolis, MN.}}
  \maketitle

\if0\anon
{
  \bigskip
  \bigskip
  \bigskip
  \begin{center}
    {\LARGE\bf Title}
\end{center}
  \medskip
} \fi

\bigskip
\begin{abstract}
The problem of detecting new signals in the presence of an unknown background is ubiquitous in scientific discoveries and is especially prominent in the physical sciences.  Most solutions proposed thus far to address the problem focus on estimating the background distribution and using that estimate to infer the signal.   
By studying the geometry of the problem, this article demonstrates that estimating the background distribution is somewhat unnecessary for inferring the signal intensity. Instead, it suffices to estimate a single parameter, referred to as \emph{the compensator}, to account for the incomplete knowledge of the background, substantially reducing the problem's complexity and enabling proper uncertainty propagation. Such a compensator is shown to govern the conservativeness of the inference, both in the proposed setup and in likelihood-based approaches.

\end{abstract}

\noindent%
\vfill

\newpage
\spacingset{1.8} 

\section{Introduction}
\label{sec:intro}
The discovery of new physics, whether it involves finding a new particle or detecting a new astrophysical phenomenon, often consists of searching for a signal from a specific source of interest in the data recorded by detectors. While the data at hand are not guaranteed to contain the signal from the targeted source, they typically collect signals from a wide variety of nuisance sources, known as \emph{the background}. Hence, the problem of signal detection depends heavily on the knowledge available about the latter.

The true data distribution, $F$, with density $f$, can be expressed as a convex combination of the distribution of the signal of interest, $F_s$, with density $f_s$, and a background distribution $F_b$ with density $f_b$, i.e.,
\begin{equation}
    \label{eq:mix_model}
    f(x;\eta) = \eta f_s(x) + (1-\eta)f_b(x);  \hspace{5pt} x \in \mathcal{X}; \hspace{5pt} \eta \in [0,1)
\end{equation}
where $\mathcal{X}\subset \mathbb{R}^d, d \ge 1$, is the search region, over which the recorded observations lie, and is assumed to be a compact set; $\eta$ is the true proportion of signal events, i.e., the relative amount of data generated by the signal of interest. To exclude the case in which the data collected from the experiments contain no background events, which is unrealistic in practical applications, it is sensible to assume $\eta$ to be bounded away from one. 

The signal detection problem can be formulated as a statistical test for the hypotheses: 
\begin{equation}
    \label{eq:eta_test}
    H_{0,\eta}: \eta = 0 \quad\text{vs}\quad H_{1,\eta}: \eta > 0
\end{equation}
in which the null hypothesis, $H_{0,\eta}$, corresponds to the situation in which the observed data consists only of background events. Thus, rejecting this hypothesis would provide statistical evidence for the presence of the signal of interest. 

In many instances, rather than recording a continuous stream of `unbinned' data, the measurements obtained consist of counts observed over a large number of bins \citep[Cf.][]{algeri_khmaladze}. In these settings, the data-generating process is usually modeled as a Poisson process with intensity function proportional to the density in \eqref{eq:mix_model} \citep[e.g.][]{bkg_modeling, discrete_profiling}.
Thus, while the testing problem in \eqref{eq:eta_test} is also relevant in the binned data scenario, we focus here on the case where the data are independent and identically distributed (i.i.d.). An extension to the binned data regime is pursued in \citet{banerjee_algeri_binned}.

In what follows, we assume the signal density $f_s$ is fully specified, which is crucial to our methodology. This assumption is realistic because physical theories extensively investigate the behavior of the source we aim to detect and can reliably model its distribution.  Extensions to situations in which $f_s$ is known only up to some free parameters are also possible and are discussed in Section \ref{sec:discussion}. 
The background density, $f_b$, however, remains unknown as the multitude of sources contributing to it and their interactions are not well understood. This renders testing \eqref{eq:eta_test} especially challenging. In particular, overestimating the background in the region where the signal is expected to occur yields overly conservative inference. More critically, underestimating the background in the signal region may lead to false discoveries. 

Most solutions proposed in the literature to address the problem rely on estimating $f_b$ before proceeding to test \eqref{eq:eta_test}. Such an estimate is obtained by considering a labeled dataset assumed not to contain any signal events, hereinafter referred to as the `background-only data', and then used in place of $f_b$ in \eqref{eq:mix_model} to test  \eqref{eq:eta_test} via a likelihood ratio test (LRT). The latter is conducted on the unlabeled dataset of interest, hereinafter referred to as the `physics data', generated by the mixture model in \eqref{eq:mix_model} and, therefore, may (if $\eta > 0$) or may not (if $\eta = 0$) contain the signal of interest. 
For example, \citet{algeri2020detecting} describes a method that models the ratio of the true and postulated background densities semiparametrically using an orthonormal expansion estimated from background-only data. 
\citet{zhang_alageri_2023} employs such an approach, along with a correction for multiple testing, to detect one or more line emissions (the signal) in high-resolution X-ray spectra under high background. 
Other methods use auxiliary events expected to behave similarly, but not exactly, as the background distribution of interest. A naive background estimate based on such data is then adequately corrected by estimating the density ratio between the true and auxiliary background distribution or via optimal transport \citep{bkg_modeling} on data collected on `control regions', i.e., regions of the support known not to contain signal events and acting as the background-only sample. The resulting estimate is then extrapolated to the `signal region' in which the signal, if present, is expected to lie. None of these methods, however, accounts for the uncertainties affecting the estimate of the background distribution when testing for the presence of the signal.

The methods proposed in \citet{safeguard}  and \cite{atlas_spurious_signal}, introduced, respectively, by members of the dark matter community and the ATLAS experiment at CERN, focus on the construction of an estimate of $f_b$ in which a signal-like component is introduced in order to `safeguard' against false discoveries. However, as demonstrated in \citet{PRD} with a rather compelling example (see also Section \ref{sec:LRT_failure}), this approach does not guarantee conservative inference and can still lead to false discoveries. Moreover, such methods fail to address the issue of model misspecification under $H_{0,\eta}$, which naturally arises due to the signal-like component introduced in the background model. As a result, the usual $\bar{\chi}^2_{01}$ approximation of the null distribution of the LRT statistic, arising when $\eta$ is tested at the boundary of its parameter space, no longer applies, making the outcome of a signal search based on such a test unreliable. 

An approach that uses only the physics dataset for testing \eqref{eq:eta_test} is the `discrete profiling method' introduced by \citet{discrete_profiling} as a part of the CMS experiment \citep{cms2012observation} at CERN. Given a predetermined pool of candidate background models, a discrete index is assigned to each of its elements, and for each of them, a profile negative log-likelihood curve is obtained with respect to the signal strength parameter. An envelope is then constructed around the corresponding minima with the goal of approximating the overall profile curve. The minimizer of this curve is reported as the final estimate of the signal strength. While avoiding the drawbacks of the `safeguard'-style methods, this method is also not exempt from limitations. For example, when considering candidate background models with varying numbers of parameters, a `$p$-value correction' is used to adjust for model complexity. The validity of such a correction, however, is only based on a heuristic argument. Moreover, it is unclear how to select the set of candidate background models from the infinite number of choices available.

This paper proposes a shift of perspective. It shows that, when a background-only dataset is available, valid inference on $\eta$ can be obtained even when the user only has access to a misspecified description of $f_b$. In particular, by studying the geometry of the problem, we demonstrate that estimating $f_b$ is somewhat unnecessary. Instead, it is sufficient to estimate a single, identifiable, scalar parameter, hereafter referred to as \emph{the compensator} (formally defined in Section \ref{sec:math_framework}), to account for the incomplete knowledge of the background. In essence, the compensator captures the deviation between the true background $f_b$ and a misspecified background density in the direction of the signal $ f_s$. Consequently, the complexity of the investigation is significantly reduced, as consistently estimating the compensator is much easier than consistently estimating $f_b$ and 
does not require the specification of a background model that can closely approximate the true background (see Section \ref{sec:inf_unbinned_wbkg}). Moreover, the uncertainties associated with estimating the compensator can be naturally accounted for when inferring $\eta$.
The compensator parameter is also shown to arise in LRT-based signal search methods and determines their degree of conservativeness under model misspecification.  

In the absence of background-only data, the compensator is no longer identifiable. Nevertheless, our geometric study lays the groundwork for a sensitivity analysis that allows us to evaluate the degree of conservativeness of the proposed inferential approach and prevent false discoveries.

The remainder of the paper is organized as follows. In Section \ref{sec:math_framework}, we describe the geometry of the problem at hand and discuss the shortcomings affecting LRT-based approaches. Section \ref{sec:inf_unbinned_wbkg} introduces the proposed estimation and testing framework for $\eta$ when a background-only sample is available to the user. Section \ref{sec:inf_wobkg} describes the proposed sensitivity analysis for the case in which a background-only dataset is not available. Section \ref{sec:data_analysis} demonstrates the effectiveness of the proposed methodology using a realistic simulation of Fermi Large Area Telescope data. Extensions are discussed in Section \ref{sec:discussion}. Technical proofs, additional figures, data, and codes are collected in the Appendix.

\section{On the geometry of the problem}
\label{sec:math_framework}
As discussed in Section \ref{sec:intro}, a description of the background density $f_b$ may only be approximately known or completely unknown. Begin by specifying a postulated background density $g$ with distribution function $G$. Assume the latter shares the same support as $F_s$ and $F_b$ and $\frac{f_s}{g}, \frac{f_b}{g} \in \Hl$. 
The density $g$ is not required to provide a close approximation of the true background density $f_b$: at least for the case in which a background-only sample is available to the user, valid inference on $\eta$ can be obtained even when $g$ is substantially different from $f_b$. The robustness of the method with respect to the choice of $g$ is demonstrated in Sections \ref{sec:inf_unbinned_wbkg} and \ref{sec:data_analysis} through practical examples.

To aid intuition, begin with the case in which $g$ is fully specified. The case where $g$ depends on unknown parameters is discussed in Section \ref{sec:inf_unbinned_wbkg}. 
Consider an orthonormal basis $\mT$ for $\Hl$ such that $\mathcal{T}=\{1,\Sd,T_1,T_2,\dots\}$ in which
\begin{equation}
\label{eqn:basis}
\begin{split}
\Sd=\frac{S}{\|S\|_{G}},&\quad S = \frac{f_s}{g} - 1, \quad \normS^2 = \intlu S^2(x)dG(x),
\end{split}
\end{equation}
and the remaining elements $\{T_j\}_{j\geq 1}$ are constructed by taking any orthonormal basis in $\Hl$, namely $\mT^*$, 
and orthonormalizing its elements with respect to $\Sd$ via Gram-Schmidt orthogonalization so that
\begin{equation*}
\begin{split}
\langle T_j, T_{j'}\rangle_{G} &= \int_{\mathcal{X}} T_j(x)  T_{j'}(x) dG(x) = \mathds{1}_{\{j=j'\}}; \quad
\langle T_j, 1\rangle_{G} = \int _{\mathcal{X}}T_j(x)dG(x) = 0\\
\langle T_j, \Sd\rangle_{G} &= \int_{\mathcal{X}} T_j(x)  \Sd(x) dG(x) = 0; \quad
\langle \Sd, 1\rangle_{G} = \int _{\mathcal{X}}\Sd(x)dG(x) = 0\\
\end{split}
\end{equation*}
where $\mathds{1}_{\{\cdot\}}$ is the indicator function. It is possible to show that (see Appendix \ref{app:completeness}) if  $\mT^*$ is complete then so is $\mT$. Hence, we can expand any function in $\Hl$ using $\mT$. 
In particular, $f_b$ can be expressed as:
\begin{equation}
    \label{eq:fb_by_g}
   f_b(x)=g(x)\biggl[ 1 + \sum\limits_{j=1}^\infty \zeta_j T_j(x) + \delta \Sd(x)\biggl];
\end{equation}
where the term in the square brackets is an orthonormal expansion of the density ratio $\frac{f_b}{g}$. The coefficients of this expansion specify as:
\begin{equation}
    \label{eq:beta_delta_def}
        \zeta_j = \left\langle \frac{f_b}{g}, T_j \right\rangle_{G} = \int_{\mathcal{X}} T_j(x)dF_b(x)\quad\text{and}\quad
        \delta = \left\langle \frac{f_b}{g}, \Sd \right\rangle_{G} = \int_{\mathcal{X}} \Sd(x)dF_b(x).
\end{equation}

The function $\Sd$ is a normalized version of the score function for the density $f$ in \eqref{eq:mix_model} evaluated at $\eta = 0$, when $f_b \equiv g$. It describes the direction from which deviations of the postulated background toward the signal model occur. By ensuring its inclusion in the basis $\mathcal{T}$, such a function allows us to explicitly oversee how such deviations affect our inference not only on $f_b$ but, most importantly, on $\eta$. 

In particular, by plugging in \eqref{eq:fb_by_g} into \eqref{eq:mix_model} and rearranging the terms, we can express $f$ as:
\begin{equation}
    \label{eq:f_by_g}
    f(x) = g(x)\left[
    1 + \sum\limits_{j=1}^\infty \tau_j T_j(x) + \theta \Sd(x)
    \right]
\end{equation}
where,
\begin{equation}
\label{eq:theta_tau_def}
    \tau_j = (1-\eta) \zeta_j = \int_{\mathcal{X}} T_j(x) dF(x), \hspace{10pt}
   \theta = \eta\normS +(1-\eta)\delta  = \int_{\mathcal{X}} \Sd(x)dF(x);
\end{equation}
consequently,
\begin{equation}
    \label{eq:eta_def}
    \eta = \frac{\theta - \delta}{\normS - \delta}.
\end{equation}
In \eqref{eq:f_by_g}, all $\tau_j$ parameters and $\theta$  are identifiable and from \eqref{eq:eta_def} it follows that, when $\delta=0$, the process of inferring $\eta$ is equivalent to that of inferring $\theta$. Whereas, when $\delta \ne 0$, $\theta$ can still be employed to infer $\eta$. The parameter $\delta$, however, now intervenes by compensating for the differences between $g$ and $f_b$. For this reason, as anticipated in Section \ref{sec:intro}, we refer to such a parameter as \emph{the compensator}.

\begin{figure}[t]
  \centering
    \includegraphics[width=\textwidth]{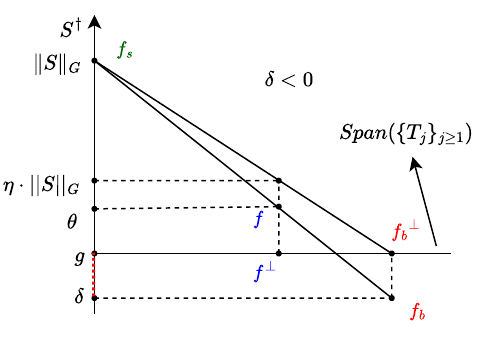}\hspace{0.5cm}
    \includegraphics[width=\textwidth]{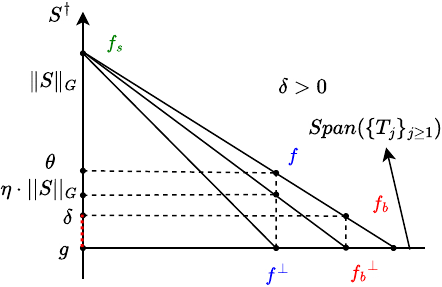}
  \caption{Relative positions of the density ratios with respect to our postulated background $g$ in the space spanned by $\mT$. Each point, labeled by a given density $h$, represents a function of the form $\frac{h}{g} - 1$. The left and the right panels represent the scenarios when $\delta$ is negative and $\delta$ is positive, respectively. The horizontal axis corresponds to the space spanned by $\{T_j\}_{j \ge 1}$, while the vertical axis represents the space spanned by $\Sd$. The compensator,  $\delta$, is the length of the projection of $f_b$ onto the vertical axis (dotted red segment), capturing the deviation between $f_b$ and $g$ in the direction of $f_s$.}
  \label{fig:func_space}
\end{figure}

Figure \ref{fig:func_space} shows the relative positions of the density ratios involved in our study with respect to $g$ in the subspace spanned by $\mT$, when $\delta<0$ (left panel) and $\delta>0$ (right panel). To ease the notation in the proposed diagrams, the position of a function of the form $\frac{h}{g}-1$ is labeled by the corresponding density $h$. For example, in each diagram, $f_s$ serves as a label for $S = \frac{f_s}{g}-1$.
The functions $f^\perp$ and $f_b^\perp$ are, respectively, the projections of $f$ and $f_b$ onto $Span(\{T_j\}_{j\ge 1})$. Note that $\delta$ is zero when either  $f_b\equiv f^\perp_b$, i.e., the ratio $\frac{f_b}{g}$ is completely spanned by the $T_j$'s or $f_b\equiv g$. In all the other cases, the compensator $\delta$ captures the sign and the magnitude of the deviation between $f_b$ and $g$ in the direction of $f_s$ (dotted red segment on the vertical axis).

Notice that, in \eqref{eq:eta_def}, we are implicitly assuming that $\delta \ne \|S\|_{G}$. This is a reasonable assumption since, geometrically, $\delta = \normS$ holds when the projection of the background $f_b$ onto the vertical axis in Figure \ref{fig:func_space} is of the same length as that of the signal $f_s$. This indicates that, in the expansion for $f_b$, the signal component $\Sd$ is as prominent as it is for the signal itself. Therefore, violations of this assumption would only occur in rather implausible scenarios.

While $\delta$ is not identifiable under \eqref{eq:f_by_g}, it is under \eqref{eq:fb_by_g}. It follows that, as discussed in Section \ref{sec:inf_unbinned_wbkg}, when a sample from $F_b$ is available, $\delta$ can be consistently estimated via \eqref{eq:fb_by_g} on such a sample and, by virtue of \eqref{eq:eta_def}, the resulting estimate enables us to both estimate $\eta$ consistently and test \eqref{eq:eta_test} while ensuring the desired probability of type I error.

On the other hand, in the absence of a background-only sample from $F_b$, if $\delta<0$, it is still possible to perform valid, yet conservative, inference on $\eta$ by relying on the estimable quantity  $\frac{\theta}{\|S\|_{G}}<\eta$. Caution must be taken, however, to avoid situations in which $\delta$ is positive, which results in $\frac{\theta}{\|S\|_{G}} > \eta$, leading to overestimating the signal proportion and an inflation of the probability of type I error. The relationship between $\frac{\theta}{\normS}$ and $\eta$ can also be examined visually by comparing $\theta$ and the position of $\eta\normS$ in Figure \ref{fig:func_space}. In particular, as $\delta$ gets closer to zero, our bias in estimating $\eta$ with $\frac{\theta}{\normS}$ decreases. A sufficient condition that ensures the negativity of $\delta$ in bump-hunting problems is identified in Section \ref{sec:inf_wobkg}; whereas, Section \ref{sec:LRT_failure} demonstrates the role of $\delta$  when relying on tests based on the LRT.

\subsection{Anti-conservativeness of LRT-based methods}
 \label{sec:LRT_failure}
As anticipated in Section \ref{sec:intro}, the first step of most likelihood-based approaches is to estimate $f_b$ on a background-only sample. Denote such an estimate with $\tilde{g}$. In order to test \eqref{eq:eta_test} using the LRT, $\tilde g$ is plugged into \eqref{eq:mix_model} in place of $f_b$; thus, instead of estimating $\eta$ via the maximum likelihood estimate (MLE) based on \eqref{eq:mix_model}, such a parameter is estimated by the MLE of ${\tilde\eta}$ in the model:
\begin{equation}
\label{eq:LRT_missp_model}
    \tilde{f}(x; \tilde\eta) = \tilde \eta f_s(x) + (1-\tilde \eta) \tilde{g}(x)\quad\text{with}\quad\tilde\eta\in[0,1).
\end{equation}
\begin{figure}[t]
  \centering
  \begin{subfigure}[b]{0.5\textwidth}
    \centering
    \includegraphics[width=\textwidth]{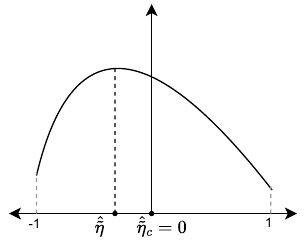}
    \caption{$\widehat{\tilde\eta} < 0, \widehat{\tilde\eta}_c = 0$}
    \label{fig:eta_hat<0}
  \end{subfigure}\hfill
  \begin{subfigure}[b]{0.5\textwidth}
    \centering
    \includegraphics[width=\textwidth]{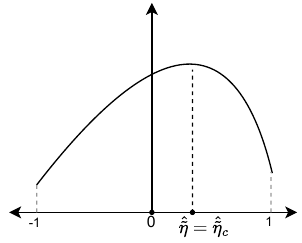}
    \caption{$\widehat{\tilde\eta} = \widehat{\tilde\eta}_c > 0$}
    \label{fig:eta_hat>0}
  \end{subfigure}
  \caption{Plots of the log-likelihood function $\ell_n$ as a function of $\tilde\eta$. The positions of the the M-estimator $\widehat{\tilde{\eta}}$ and, the MLE $\widehat{\tilde\eta}_c$ are shown for the cases when is $\widehat{\tilde{\eta}}$ negative (left panel) and when $\widehat{\tilde{\eta}}$ is positive (right panel).}
  \label{fig:eta_hat_eta_hat_c}
\end{figure}
Note that in \eqref{eq:eta_test}, the value of $\eta$ under the null hypothesis lies on the boundary of its parameter space. While standard likelihood-based inference assumes the true value of the parameter to be an interior point of the parameter space \citep[e.g.,][p.~444]{lehmann1998theory}, \citet{chernoff1954distribution} and \citet{self_and_liang}, among others, studied the asymptotic distribution of the MLE and the LRT statistic when the parameter value lies on the boundary, and no model misspecification arises.
However, to cover situations in which $\tilde{g}\not \equiv f_b$, as is the case, at least, for the safeguard approach  \citep{safeguard} and the spurious signal method \citep{atlas_spurious_signal} (see Section \ref{sec:intro}), it is necessary to assess how the MLE for $\tilde\eta$ in \eqref{eq:LRT_missp_model} relates to $\eta$ in \eqref{eq:mix_model} and its impact on the asymptotic behavior of the LRT.

To facilitate the exposition, begin by extending the parameter space
from $[0,1)$ to $[-1, 1]$, and define the following estimators:
\begin{equation}
    \label{eq:eta_hat_c_eta_hat_def}
    \widehat{\tilde\eta} = \arg\max_{\tilde\eta \in [-1,1]} \ell_n(\tilde\eta) \quad\text{and}\quad\hspace{10pt}
    \widehat{\tilde\eta}_c = \arg\max_{\tilde\eta \in [0,1)} \ell_n(\tilde\eta)
\end{equation}
where $\ell_n(\tilde\eta) = \sum_{i=1}^n \log\tilde{f}(X_i; \tilde\eta)$ is the log-likelihood function based on \eqref{eq:LRT_missp_model} computed on the physics sample. Formally, the latter consists of observations of $n$ i.i.d. random variables  $X_1, X_2, \cdots, X_n  \sim F$. Note that, $\widehat{\tilde{\eta}}_c$ is the MLE for $\tilde\eta$, whereas, $\widehat{\tilde\eta}$ is an M-estimator but not an MLE. This is because $\tilde f(\cdot; \tilde\eta)$ may not be non-negative for some negative values of $\tilde\eta$. Nonetheless, the theory of M-estimation allows us to establish the asymptotic limit of $\widehat{\tilde\eta}$ and, in turn, that of $\widehat{\tilde\eta}_c$ by exploiting their relationship.

\begin{figure}[t]
  \centering
  \begin{subfigure}[b]{0.45 \textwidth}
    \centering
    \includegraphics[width=0.7\textwidth]{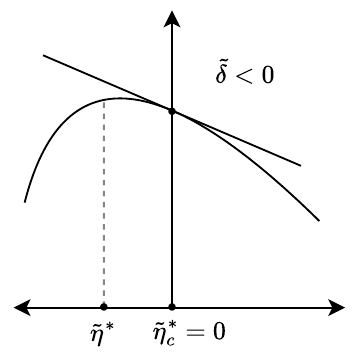}
    \caption{$\mathbb{E}_{F_b}[\log \tilde f(X;\tilde\eta)]$ when  $\tilde\delta < 0$}
    \label{fig:tilde_del<0}
  \end{subfigure}\hfill
  \begin{subfigure}[b]{0.45\textwidth}
    \centering
    \includegraphics[width=0.75\textwidth]{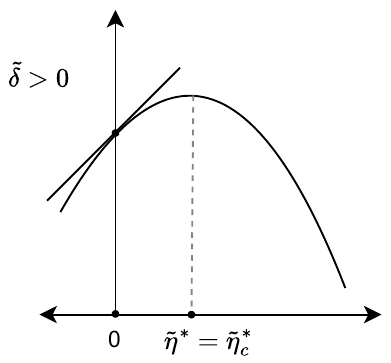}
    \caption{$\mathbb{E}_{F_b}[\log \tilde f(X;\tilde\eta)]$ when  $\tilde\delta > 0$}
    \label{fig:tilde_del>0}
  \end{subfigure}
  \caption{Plots of the expected log-likelihood function $\mathbb{E}_{F_b}[\log \tilde f(X;\tilde\eta)]$ as a function of $\tilde\eta$. The left and the right panel show the positions of the asymptotic limits $\tilde\eta^*$ and $\tilde\eta^*_c$ when $\tilde\delta$ is negative, and $\tilde\delta$ is positive, respectively.}
  \label{fig:E_fb_log}
\end{figure}
Since
\begin{equation}
    \label{eq:concavity}
    \frac{\partial^2}{\partial \tilde\eta^2} \log \tilde{f}(x;\tilde\eta) = -\left(\frac{\tilde{g}(x) - f_s(x)}{\tilde{f}(x;\tilde\eta)}\right)^2 < 0,
\end{equation}
the log-likelihood $\ell_n$ is a strictly concave function of $\tilde\eta$ implying that $\ell_n(\tilde\eta)$ has a unique global maximum. Hence, when the maximizer of $\ell_n(\tilde\eta)$ on $[-1,1]$, i.e. $\widehat{\tilde \eta}$, is negative (see Figure \ref{fig:eta_hat<0}), $\ell_n$ must be strictly decreasing on $[0,1)$ leading to $\widehat{\tilde \eta}_c = 0$. On the other hand, if $\widehat{\tilde\eta} > 0$ (see Figure \ref{fig:eta_hat>0}), the maximizer of $\ell_n$ on $[-1,1]$ and that on $[0,1)$ coincide, hence:
\begin{equation}
    \label{eq:eta_hat_c_eta_hat}
    \widehat{\tilde\eta}_c = \widehat{\tilde\eta}\cdot\mathds{1}_{\{\widehat{\tilde\eta} > 0\}}.
\end{equation}

We assume, throughout the paper, that the densities $f_s$ and $f_b$ are continuous in $x$. Note that, under $H_{0,\eta}$, $X_1, X_2, \cdots, X_n\sim F_b$. Then, by \citet[Theorem 5.9]{Vaart_1998}, we get $\widehat{\tilde\eta}$ converging to
the minimizer of the Kullback-Leibler (KL) divergence between $F_b$ and $\tilde{F}$, hereafter denoted by ${\tilde\eta}^*$, and such that:
\[
{\tilde\eta}^*=
\arg\max_{{\tilde\eta} \in [-1,1]}\mathbb{E}_{F_b}\left[\log\tilde f(X;{\tilde\eta})\right]
\]
where $\mathbb{E}_{F_b}[\cdot]$ denotes the expectation taken with respect to $F_b$.
Since  $\widehat{\tilde\eta} \stackrel{\mathbb{P}}\longrightarrow {\tilde\eta}^*$ under $H_{0,\eta}$ and since $x\mathds{1}_{\{x>0\}}$ is continuous, from \eqref{eq:eta_hat_c_eta_hat} it follows that,
\begin{equation}
    \label{eq:eta_hat_c_limit}
    \widehat{\tilde\eta}_c \stackrel{\mathbb{P}}\longrightarrow {\tilde\eta}^*\mathds{1}_{\{{\tilde\eta}^* > 0\}} = {\tilde\eta}^*_c.
\end{equation}
Moreover, since  $\log \tilde{f}$ is continuous with respect to $\tilde\eta$ and $x$ on the compact sets $[-1,1]$ and $\mathcal{X}$, respectively, it is bounded, and hence the dominated convergence theorem allows differentiation with respect to ${\tilde\eta}$ under the integral with respect to $F_b$. Thus, we have:
\begin{equation}
\label{eq:LRT_exp_score}
\begin{aligned}
    \left. \frac{\partial}{\partial\tilde\eta} \mathbb{E}_{F_b}\left[\log \tilde f(X; {\tilde\eta})\right]\right|_{{\tilde\eta} = 0} &= 
    \int_{\mathcal{X}} \left. \frac{\partial}{\partial \tilde\eta} \log \tilde{f}(x;{\tilde\eta}) dF_b(x) \right |_{{\tilde\eta} = 0} 
    = \int_{\mathcal{X}} \left(\frac{f_s(x)}{\tilde{g}(x)}-1\right)dF_b(x).
\end{aligned}
\end{equation}
Observe that the integrand in the last equality is analogous to $S$ in \eqref{eqn:basis} with  $\tilde{g}$ in place of $g$. We can, therefore, define the analogues to $S$, $\Sd$ and $\delta$ for the present setting as:
\[
\tilde{S}(x) = \frac{f_s(x)}{\tilde{g}(x)} - 1,
\quad
\tilde{\Sd}(x) = \frac{\tilde S(x)}{\|\tilde{S}\|_{\tilde{G}}},\quad \tilde{\delta} = \int_{\mathcal{X}}\tilde{\Sd}(x)dF_b(x),
\]
and rewrite \eqref{eq:LRT_exp_score} as
\[
\left. \frac{\partial}{\partial{\tilde\eta}} \mathbb{E}_{F_b}\left[\log \tilde f(X; {\tilde\eta})\right]\right|_{{\tilde\eta} = 0} = \int_{\mathcal{X}} \tilde{S}(x)dF_b(x) = 
\|\tilde{S}\|_{\tilde{G}} \tilde{\delta}.
\]
The above expression implies that the sign of the slope of $\mathbb{E}_{F_b}\left[\log \tilde f(X; {\tilde\eta})\right]$ at ${\tilde\eta} = 0$ is solely determined by the sign of $\tilde\delta$. Moreover, from \eqref{eq:concavity}, it follows that $\mathbb{E}_{F_b}\left[\log \tilde{f}(X;{\tilde\eta})\right]$ is also strictly concave in ${\tilde\eta}$ and, therefore, has a unique maximum. This tells us that, when $\tilde\delta$ is negative, as shown in Figure \ref{fig:tilde_del<0}, the unconstrained maximizer ${\tilde\eta}^*$ must also be negative and, from \eqref{eq:eta_hat_c_limit}, we have ${\tilde\eta}^*_c = 0$. Thus, in this scenario, $\widehat{\tilde\eta}_c$ is still a consistent estimator of $\eta$ under $H_{0,\eta}$. However, when $\tilde\delta > 0$ (see Figure \ref{fig:tilde_del>0}) the maximizer ${\tilde\eta}^*$ is strictly positive and coincides with ${\tilde\eta}^*_c$. Hence, in this case, $\widehat{\tilde\eta}_c$ is no longer consistent for ${\eta} = 0$ and has a positive limit. This indicates that a positive value of $\tilde\delta$ can result in an inflated estimate of the true signal intensity $\eta$ under $H_{0,\eta}$.

Let us now consider the LRT statistic for testing \eqref{eq:eta_test} with $\tilde\eta$ in place of $\eta$, i.e.,
\begin{equation}
    \label{eq:LRT_stat}
    -2[ \ell_n(0)- \ell_n(\widehat{\tilde\eta}_c)].
\end{equation}
\begin{figure}[t]
  \centering
    \includegraphics[width=0.48\textwidth]{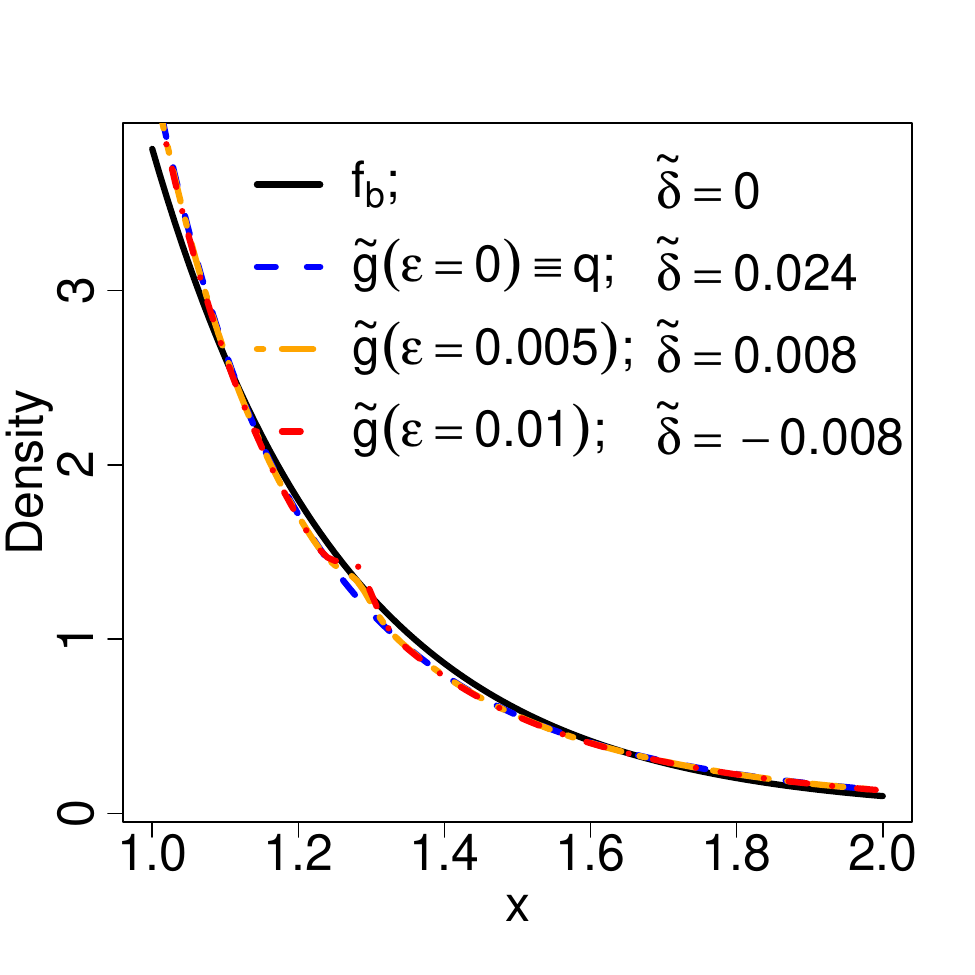}
    \centering
    \includegraphics[width=0.48\textwidth]{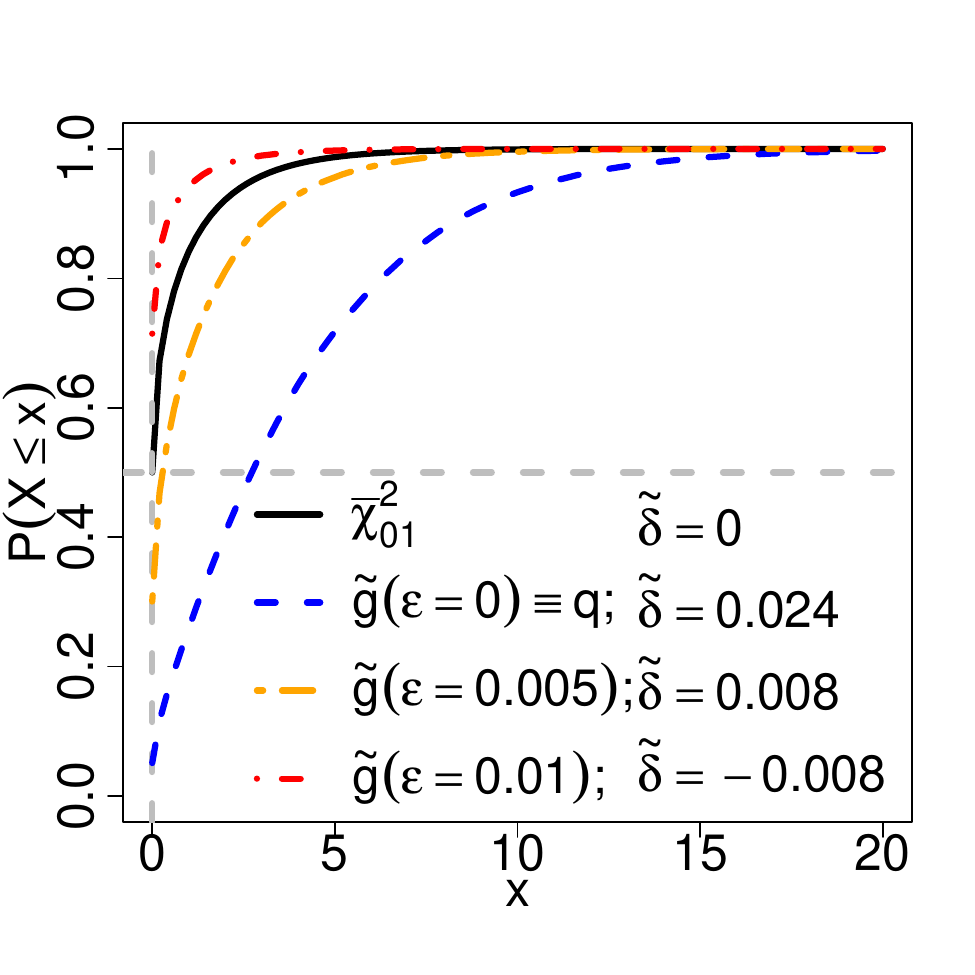}
    \caption{
  Left Panel: Graphs of the true background density $f_b$ and the misspecified background density $\tilde g$ for increasing values of $\varepsilon$, along with the corresponding values for the compensator $\tilde\delta$. Right panel: Comparing the reference distribution $\bar\chi^2_{01}$ (black solid line) with the simulated null distributions of the LRT statistic for $\varepsilon = 0$ (blue dashed line), $\varepsilon =0.005$ (orange two-dashed line), and  $\varepsilon =0.01$ (red dot-dashed line).}
  \label{fig:LRT_missp}
\end{figure}
If $\tilde{g} \equiv f_b$, the limiting null distribution of \eqref{eq:LRT_stat} is the usual $\bar\chi^2_{01} = \frac{1}{2}\chi^2_0 + \frac{1}{2}\chi^2_1$ distribution. However, for methods such as the safeguard and spurious signal approaches, in which the background is systematically misspecified due to the introduction of the signal-like bump into the background density, the asymptotic null distribution of the LRT is no longer $\bar\chi^2_{01}$. To illustrate this phenomenon with a simple example, let us consider a gamma-distributed background, truncated on the interval [1,2], i.e.,
\begin{equation}
    \label{eq:LRT_sim_fb}
    f_b(x) \propto \exp(-3.3\cdot x)x^{-1/2},\quad x\in[1,2].
\end{equation}
The misspecified background model in \eqref{eq:LRT_missp_model}, inclusive of a spurious signal is:
\begin{equation}
\label{eq:LRT_sim_missp}
\begin{aligned}
    & \tilde{g}(x; \varepsilon) = \varepsilon f_s(x) + (1-\varepsilon) q(x),\quad\text{with} \quad q(x) \propto x^{-5}, \quad  f_s(x) \propto \exp\left\{-\frac{(x-1.28)^2}{0.0008}\right\},
\end{aligned}
\end{equation}
and $x\in[1,2]$. The density $q$ serves as a misspecified description of $f_b$, and the parameter $\varepsilon$ in $\tilde g$ controls the magnitude of the signal-like component being injected. 

A simulation involving 100,000 Monte Carlo replicates, each of size $n = 5000$, is performed to derive the distribution of the LRT statistic in \eqref{eq:LRT_stat} under $F_b$. The right panel of Figure \ref{fig:LRT_missp} compares the $\bar{\chi}^2_{01}$ distribution (black solid line) with the simulated distribution of the LRT for different values of the parameter $\varepsilon$. As shown on the left panel of the same figure,  $\tilde g$ exhibits a visible bump for $\varepsilon = 0.01$ (red dot-dashed line) while for $\varepsilon = 0.005$ (orange two-dashed line) the signal-like term is rather faint. The corresponding LRT distributions (the orange two-dashed line and the red dot-dashed line on the right panel) show clear deviation from the $\bar{\chi}^2_{01}$ distribution (black solid line). 
For comparison, the simulated distribution of the LRT for $\tilde g(\cdot; \varepsilon = 0) \equiv  q$ (blue dashed line) is also plotted. 
For $\varepsilon = 0.01$, $\tilde{\delta}$ is negative, and the LRT has a distribution that is stochastically smaller (red dot-dashed line) than the $\bar{\chi}^2_{01}$ distribution, leading to overly conservative inference. More importantly, even though $\tilde{g}(\cdot; \varepsilon = 0)$ and $\tilde{g}(\cdot; \varepsilon = 0.005)$ appear to  approximate $f_b$ well, the corresponding LRT statistics have distributions stochastically larger than the $\bar{\chi}^2_{01}$ distribution. In these cases, the compensator $\tilde{\delta}$ is positive, and inference based on the $\bar{\chi}^2_{01}$ approximation unavoidably leads to false discoveries. 
 
\section{Inference when a background-only sample is given}
\label{sec:inf_unbinned_wbkg}
Although LRT-based approaches can lead to anti-conservative inference, when the user has access to both background-only and physics data, valid inference on $\eta$ can be performed on the basis of the framework introduced in Section \ref{sec:math_framework}.

Begin with the case in which the proposal background $g$ is fully specified. Formally, the background-only data are realizations of $m$ i.i.d. random variables $Y_1, Y_2, \cdots, Y_m \sim F_b$ and recall that the physics data consist  of i.i.d. observations of $n$ random variables  $X_1, X_2, \cdots, X_n  \sim F$. When both datasets are available, the parameters  $\theta$ and $\delta$ are estimable. In particular, by replacing $F$ and $F_b$ in \eqref{eq:beta_delta_def} and \eqref{eq:theta_tau_def} with their empirical counterparts 
$\mathbb{F}^{(n)}(x) = \frac{\sum_{i=1}^n \mathds{1}_{\{X_i \le x\}}}{n}$ and $\mathbb{F}_{b}^{(m)}(x) = \frac{\sum_{i=1}^m \mathds{1}_{\{Y_i \le x\}}}{m}$,
we obtain:
\begin{equation}
    \label{eq:theta_hat_delta_hat_def}
    \widehat{\theta} = \intlu \Sd(x) d\mathbb{F}^{(n)}(x) = \frac{1}{n} \sum\limits_{i=1}^n \Sd(X_i)\quad\text{and}\quad
\widehat{\delta} = \intlu \Sd(x) d\mathbb{F}_b^{(m)}(x) = \frac{1}{m} \sum\limits_{i=1}^m \Sd(Y_i).
\end{equation}
Plugging in $\widehat{\theta}$ and $\widehat{\delta}$  in place of $\theta$ and $\delta$ into \eqref{eq:eta_def} leads to the following estimator of $\eta$:
\begin{equation}
    \label{eq:eta_hat_def}
    \widehat{\eta} = \frac{\widehat{\theta} - \widehat{\delta} }{\normS - \widehat{\delta}}.
\end{equation}

To account for the uncertainties associated with estimating $\delta$ on the background-only sample, while inferring $\eta$ on the physics sample, it is sensible to assume that
\begin{equation}
    \label{eq:m_n_lim} 
     \frac{n}{m+n} \to \pi \in (0,1) \quad\text{as $m ,n \to \infty$}.
\end{equation}
Such an assumption reflects the realistic scenario in which the size of the two samples is comparably large, as opposed to the case in which $\pi$ is assumed to be zero to avoid dealing with any propagation of the uncertainties on the quantities estimated on the background-only data. 

When the above assumption holds, the asymptotic distribution of $\widehat{\eta}$ is given by the following proposition:
\begin{prop}
    \label{prop:eta_unb_wbkg}
    Under \eqref{eq:m_n_lim}, 
    \begin{equation}
    \label{eq:prop_eta_unb_wbkg_conv}
        \sqrt{\frac{mn}{m+n}}(\widehat{\eta} - \eta) \dconv \mathcal{N}(0,\sigma_\eta^2)
    \end{equation}
    where
    \[
    \sigma_{\eta}^2 = (1-\pi)\frac{\sigma_{\theta}^2}{(\normS - \delta)^2} + \pi\frac{\sigma_{{\delta}}^2(\theta - \normS)^2}{(\normS - \delta)^4}
    \]
   with $\sigma_{{\theta}}^2 = \int_{\mathcal{X}} ({\Sd}(x))^2dF(x) - \theta^2$ and 
    $\sigma_{{\delta}}^2 = \int_{\mathcal{X}} (\Sd(x))^2dF_b(x) - \delta^2$.
\end{prop}
The proof is provided in Appendix \ref{app:prop_eta_unb_wbkg_proof}.  

Proposition \ref{prop:eta_unb_wbkg} offers all the elements needed to test the hypotheses in \eqref{eq:eta_test}. Specifically, from the definitions of $\widehat\theta$ and $\widehat\delta$ in \eqref{eq:theta_hat_delta_hat_def}, it directly follows that, $\widehat\theta \pconv \theta$ and 
$\widehat\delta \pconv \delta$. Moreover, it is easy to show that, 
\[
\widehat\sigma_{\theta}^2 = \frac{1}{n}\sum_{i=1}^n \left({\Sd}(X_i)\right)^2 - \widehat\theta^2 
\pconv
\sigma_{\theta}^2
\quad \text{and} \quad
\widehat\sigma_{\delta}^2 = \frac{1}{m}\sum_{i=1}^m \left({\Sd}(Y_i)\right)^2 - \widehat\delta^2 
\pconv
\sigma_{\delta}^2;
\]
hence, 
\[
\widehat\sigma_{\eta}^2 = \frac{(1-\widehat\pi)\widehat\sigma_{\theta}^2}{(\normS - \widehat\delta)^2} + \frac{\widehat\pi\widehat\sigma_{\delta}^2(\widehat\theta - \normS)^2}{(\normS - \widehat\delta)^4} \stackrel{\mP}\longrightarrow \sigma_\eta^2
\]
where $\widehat\pi = \frac{n}{m+n}$. Therefore, a statistic for testing \eqref{eq:eta_test} is given by:
\begin{equation}
    \label{eq:eta_test_stat_unb_wbkg_b_knw}
    \mathcal{Z}_1 = \sqrt{\frac{mn}{m+n}}\cdot\frac{\widehat\eta}{\widehat\sigma_{\eta}} \stackrel{d}\longrightarrow \mathcal{N}(0,1) \quad\text{under }H_{0,\eta}.
\end{equation}
Letting $\mathcal{Z}_1^{obs}$ be its value observed on the data, an asymptotic $p$-value can then be computed as $1-\Phi(\mathcal{Z}_1^{obs})$ where $\Phi(\cdot)$ is the standard normal CDF. 

Notice that the above test does not depend on the choice of the basis $\mT^*$ introduced in Section \ref{sec:math_framework}. The latter is a mathematical construction that enables us to unveil the existence and role of the compensator $\delta$, but is unnecessary for inferring $\eta$. 
Moreover, it is worth emphasizing that,  since the true background distribution is unknown,  tests for \eqref{eq:eta_test} cannot be conducted via simulations. Hence, testing procedures based on (valid) asymptotics are especially valuable for the problem at hand. 

Let us now consider the case in which the proposal background density $g$ depends on an unknown parameter $\beta\in \mB \subset \mathbb{R}^p$. Denote such density with $g_\beta$ and let its distribution be $G_\beta$. Notice that, in this setting, the function $S$, and all related quantities, also depend on $\beta$ through $g_{\beta}$. Let $S_{\beta}$, $\|S_\bt\|_{G_{\beta}}^2$, $\Sd_\beta$, $\theta_\beta$, and $\delta_\bt$ be  the counterparts of $S$, $\|S\|_{G}^2$, $\Sd$, $\theta$, and $\delta$, respectively,  defined  as in \eqref{eqn:basis}, \eqref{eq:beta_delta_def}, and \eqref{eq:theta_tau_def}, with $g_{\beta}$ in place of $g$.

Let $\widehat\beta_m$ be the MLE of $\beta$ obtained on the background-only sample, i.e., 
\begin{equation}
    \label{eq:bth_def_wbkg_unb}
    \widehat\beta_m = \arg\max_{\beta \in \mathcal{B}} \sum_{i=1}^m \log g_\bt(Y_i);
\end{equation}
the parameters $\theta_{\bt}$, $\delta_\bt$, and $\eta$ can be estimated as in \eqref{eq:theta_hat_delta_hat_def}-\eqref{eq:eta_hat_def}, but with $\widehat\beta_m$ in place of $\beta$, i.e.,
\begin{equation}
    \label{eq:eta_hat_def_bt_est}
    \widehat\theta_{\bth_m} = \frac{1}{n}\sum_{i=1}^n \Sd_{\bth_m}(X_i),\quad
\widehat\delta_{\bth_m} = \frac{1}{m}\sum_{i=1}^m \Sd_{\bth_m}(Y_i),\quad\text{and} \quad
\widehat\eta_{\bth_m} = \frac{\widehat\theta_{\bth_m} - \widehat\delta_{\bth_m}}{\|S_{\bth_m}\|_{G_{\bth_m}} - \widehat\delta_{\bth_m}}.
\end{equation}

Let $\bts$ be the minimizer of the KL divergence between $F_b$ and $G_\bt$, then $\bth_m \pconv \bts$ if the following conditions hold (cf. \citealt[p~.46]{Vaart_1998}):

\begin{enumerate}[label=\textbf{(A\arabic*)}, leftmargin=2em]
    \item For any $\bt \in \mathcal{B}$, the maps $x \mapsto g_\bt(x)$ and
    $x \mapsto \frac{\partial^i g_\bt(x)}{\partial \bt^i}$ for $i = 1,2$ are continuous.\label{assump:gb_cont_x}
    \item For any $x \in \mathcal{X}$, the maps $\bt \mapsto g_\bt(x)$
    and $\bt \mapsto \frac{\partial^i g_\bt(x)}{\partial \bt^i}$ for $i = 1,2$ are continuous.\label{assump:gb_cont_bt}
    \item The parameter space $\mB \subset \Rp$ is compact.\label{assump:mB_compact}
    \item The point $\bts$ is in the interior of the parameter space $\mB$.\label{assump:bts_interior}
\end{enumerate}

Let $$\sigma_{\bts, \tht}^2 = \int_{\mathcal{X}} ({\Sd}_{\bts}(x))^2dF(x) - \theta_{\beta^*}^2\quad\text{and}\quad\sigma_{\bts, \del}^2 = \int_{\mathcal{X}} ({\Sd}_{\bts}(x)^2)dF_b(x) - \delta_{\beta^*}^2.$$ The asymptotic distribution of $\widehat\eta_{\bth_m}$ is given by the following proposition.
\begin{prop}
    \label{prop:eta_unb_wbkg_beta_est}
    If assumptions \ref{assump:gb_cont_x} - \ref{assump:bts_interior} hold, then under \eqref{eq:m_n_lim},
    \begin{equation}
    \label{eq:prop_eta_unb_wbkg_beta_est_conv}
        \sqrt{\frac{mn}{m+n}}(\widehat{\eta}_{\bth_m} - \eta) \dconv \mathcal{N}(0,\sigma_{\bts, \eta}^2) \hspace{5pt}
    \end{equation}
    where  $\sigma_{\bts, \eta}^2 =  (1-\pi)\sigma_{\bts,\theta}^2 A_{\bts}^2 + 
        \pi
        \Bigg[\sigma_{\bts,\del}^2 B_{\bts}^2 +
        (\Gamma_{\bts}^TJ_{\bts}^{-1}V_{\bts}J_{\bts}^{-1}\Gamma_{\bts}) 
        + 2B_{\bts}
        \Gamma_{\bts}^TJ_{\bts}^{-1} C_{\bts}\Bigg] $, with   
    $A_{\bts},B_{\bts}\in  \R$, $\Gamma_{\bts}, C_{\bts} \in \Rp$, and $J_{\bts}$, $V_{\bts}$ are $p \times p$ matrices,  depending on $\bts$.
\end{prop}
The explicit expressions for $A_{\bts}$, $B_{\bts}$, $J_{\bts}$, $V_{\bts}$, $\Gamma_{\bts}$ and $C_{\bts}$, along with the proof of Proposition \ref{prop:eta_unb_wbkg_beta_est}, are given in Appendix \ref{app:prop_eta_unb_wbkg_beta_est_proof}.
Notice that the asymptotic distribution of $\widehat\eta_{\bth_m}$ differ from that of $\widehat\eta$ in Proposition \ref{prop:eta_unb_wbkg} in the variance, which accounts for the uncertainty associated with estimating $\bt$ on the background-only sample. 

Similar to the case in which $g$ is fully specified, it is possible to show (see Appendix \ref{app:sig_bts_eta_est}) that there exists a consistent estimator  $\widehat\sigma_{\bth_m, \eta}^2$  for $\sigma_{\bts, \eta}^2$. 
\begin{figure}[t]
    \centering
    \includegraphics[width=0.48\linewidth]{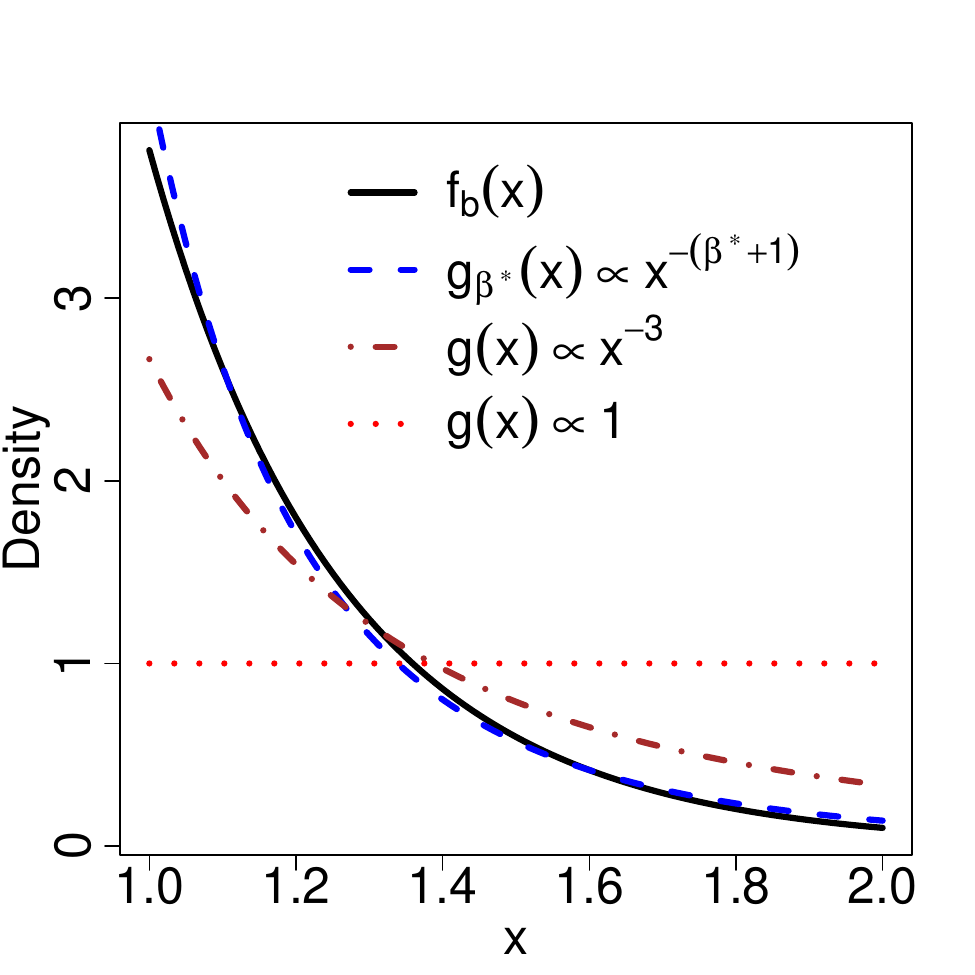}
    \includegraphics[width=0.48\textwidth]{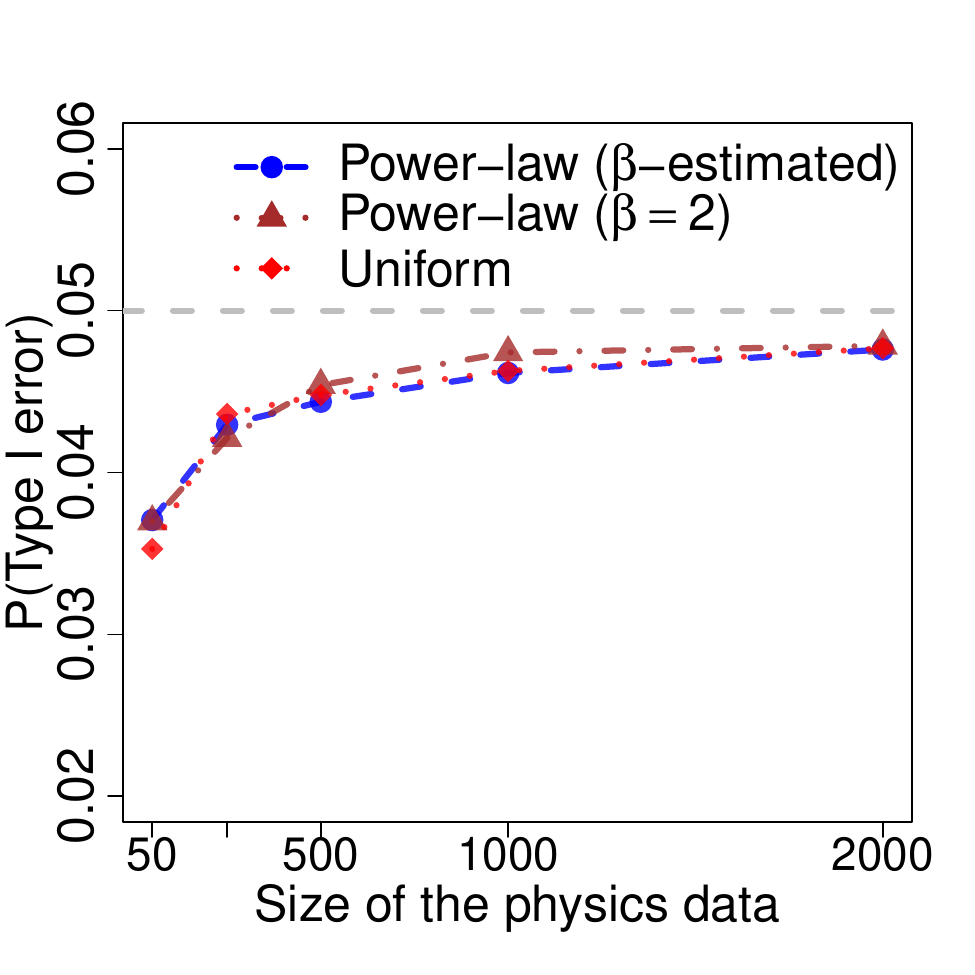}
    \caption{
    Left panel: graphs of the true background density $f_b$ as in \eqref{eq:LRT_sim_fb} (solid black line), the power-law proposal background $g_\bt$ in \eqref{eq:sim_setup_gb} for $\bt = \bts$ (blue dashed line), $\bt = 2$ (brown dot-dashed line) and a uniform proposal background (red dotted line). Right panel: Simulated probabilities of type I error.}
    \label{fig:wbkg_sim_setup}
\end{figure}
Thus, when $g_\beta$ is estimated via MLE, a test statistic for 
\eqref{eq:eta_test} is:
\begin{equation}
    \label{eq:eta_test_stat_unb_wbkg_b_est}
    \mathcal{Z}_2 = \sqrt{\frac{mn}{m+n}}\frac{\widehat\eta_{\bth_m}}{\widehat\sigma_{\bth_m,\eta}}
\end{equation}
and, letting $\mZ_2^{obs}$ be its value observed, an asymptotic $p$-value is, once again,  
$1 - \Phi(\mZ_2^{obs})$.

The tests proposed in \eqref{eq:eta_test_stat_unb_wbkg_b_knw} and \eqref{eq:eta_test_stat_unb_wbkg_b_est} are based on the expressions for $\theta$, $\delta$ and $\eta$ given by \eqref{eq:beta_delta_def}, \eqref{eq:theta_tau_def} and \eqref{eq:eta_def} respectively. Since those hold regardless of the dimensionality of $\mathcal{X}$, all  tests proposed in this section are valid for both a univariate and a multivariate setting.

\begin{figure}[t]
    \centering
    \includegraphics[width=0.33\linewidth]{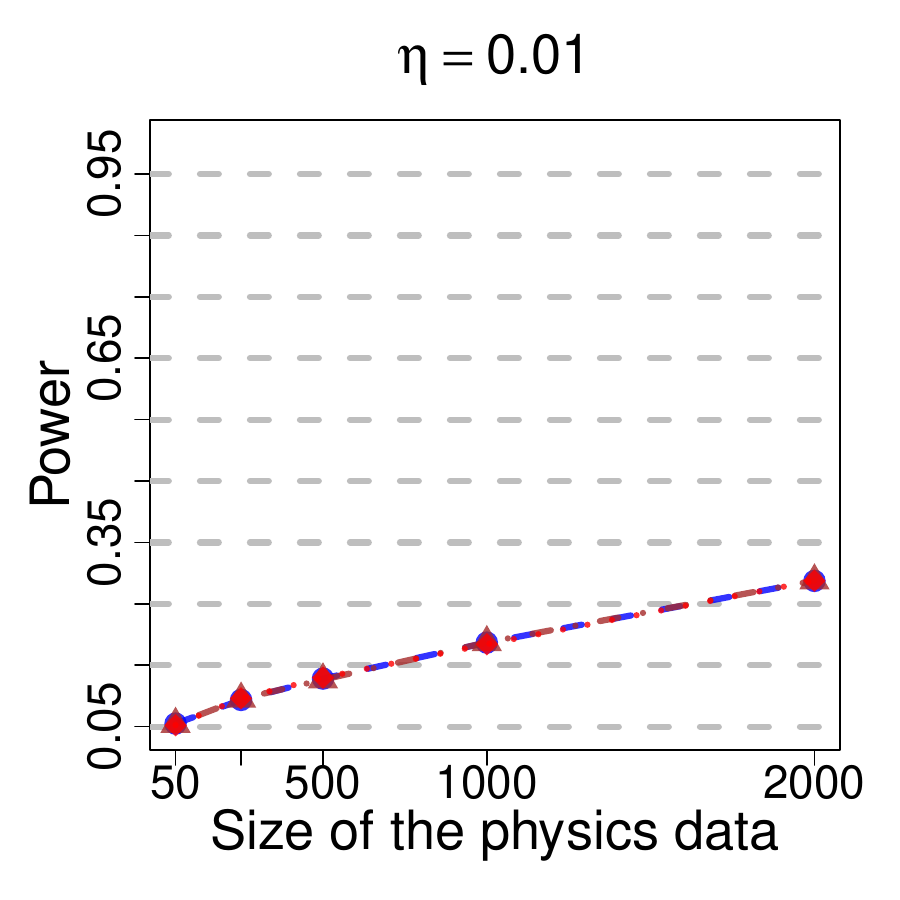}%
    \hfill%
    \includegraphics[width=0.33\linewidth]{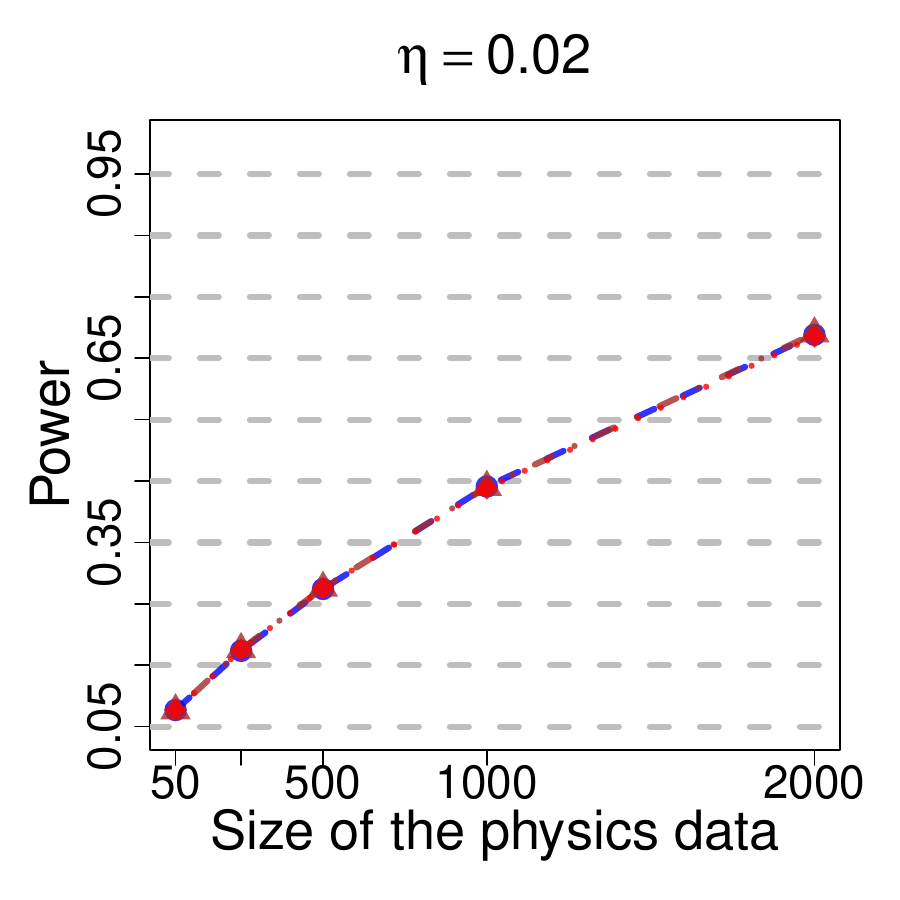}%
    \hfill%
    \includegraphics[width=0.33\linewidth]{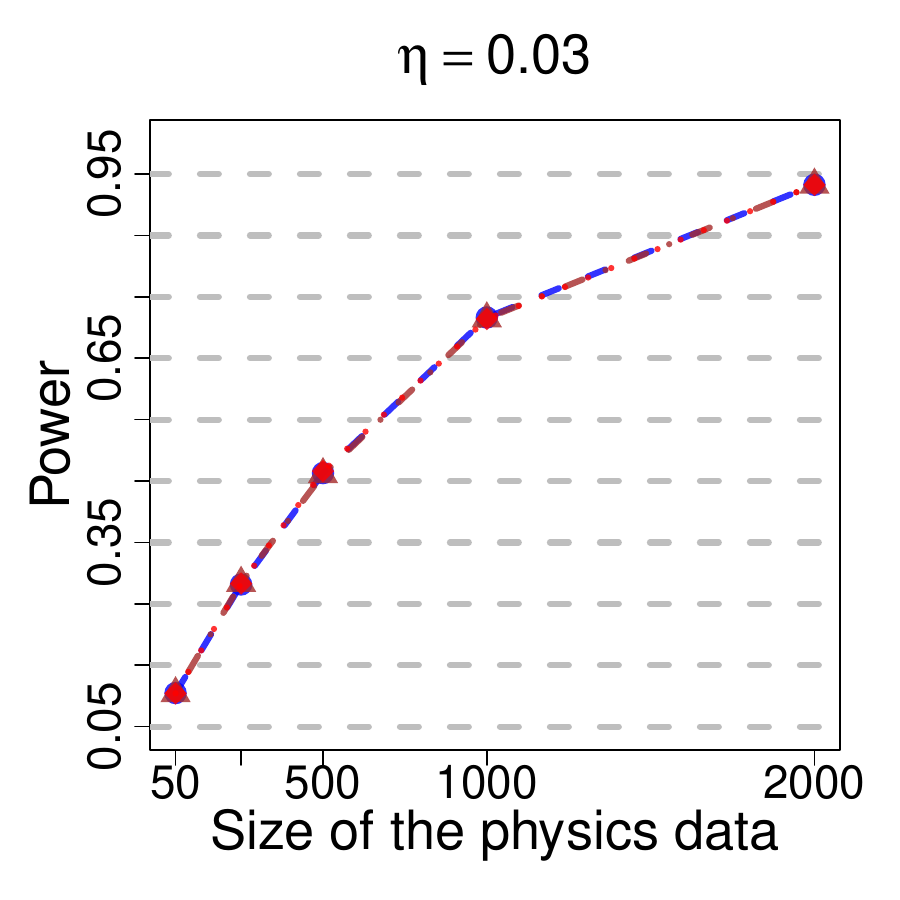}

    \begin{subfigure}[b]{\linewidth}
        \centering
        \includegraphics[width=0.9\linewidth,height=1cm]{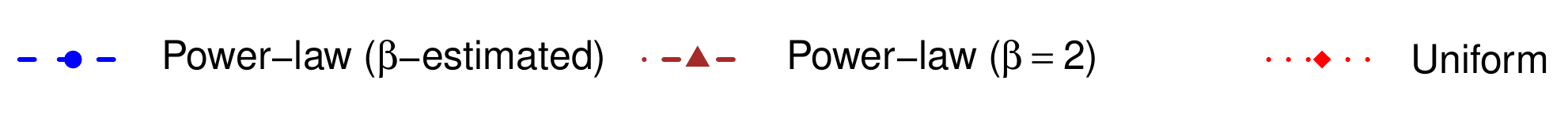}
    \end{subfigure}

    \caption{Power for various signal intensities:  $\eta = 0.01$ (left panel), $\eta = 0.02$ (central panel) and $\eta = 0.03$ (right panel); and for different choices of the postulated background: the power-law proposal background $g_\bt$ in \eqref{eq:sim_setup_gb} with $\bt = 2$ (brown dot-dashed line), the same power-law with $\bt$-estimated via MLE (blue dashed line), and a uniform density (red dotted line). The nominal significance level is 5\%.}
  \label{fig:power_curve}
\end{figure}

\textbf{Numerical example.} Let us now investigate, with an example, the statistical properties of the test statistics in 
\eqref{eq:eta_test_stat_unb_wbkg_b_knw} and \eqref{eq:eta_test_stat_unb_wbkg_b_est}, their robustness with respect to the choice of $g$, and the effect of estimating the parameter $\beta$ when dealing with a parametric $g_\bt$.

Consider a physics sample generated from \eqref{eq:mix_model} with $f_b$ and $f_s$ as in \eqref{eq:LRT_sim_fb}-\eqref{eq:LRT_sim_missp}, and choose $\eta = 0.01, 0.02$ and 0.03. The sample size is varied so that $n = 50, 250, 500, 1000$, and $2000$. For each physics sample of size $n$ considered, a background-only sample of size $m=2n$ is generated from \eqref{eq:LRT_sim_fb}. 
Letting $5\%$ be the nominal significance level, the probability of type I error and the power of the proposed procedure are simulated for three different choices of the proposal background distribution with varying degrees of departure from the true background density (see Figure \ref{fig:wbkg_sim_setup}, left panel).

First, consider a Pareto type I (or power-law) density with unknown slope, i.e.,
\begin{equation}
    \label{eq:sim_setup_gb}
    g_\bt(x) \propto x^{-(\beta + 1)}, \quad x\in [1,2]
\end{equation}
 with $\bt$ estimated via MLE on the background-only data. Since the analysis is conducted considering different sample sizes, to visualize the corresponding density, on the left panel of Figure \ref{fig:wbkg_sim_setup}, we report the graph of $g_{\bt^*}$ with $\bt^* = 3.92$ (blue dashed line); $\bts$ being the minimizer of the KL divergence between \eqref{eq:sim_setup_gb} and $f_b$ in \eqref{eq:LRT_sim_fb}  (solid black line). Next, we consider $g_\bt$ as in \eqref{eq:sim_setup_gb} but with $\beta = 2$ (brown dot-dashed line). Although such a density also decays smoothly, it deviates substantially from $f_b$. Finally, we choose $g$  to be the uniform density (red dotted line) which is entirely different from the true background.

The simulated probability of type I error for each choice of $g$ is shown on the right panel of Figure \ref{fig:wbkg_sim_setup}. The power curves for varying sample sizes and for different values of $\eta$ are given in Figure \ref{fig:power_curve}. In all cases, the simulation has been conducted using $100,000$ Monte Carlo replicates.  Figure  \ref{fig:wbkg_sim_setup} shows that, regardless of the choice of $g$ (or $g_\bt$), the simulated type I error probabilities are close to the desired significance level 0.05 and never exceed it. 
Figure \ref{fig:power_curve}  illustrates that, as expected, the simulated power increases as the physics data size and the signal intensity increase. Most importantly, the power curves corresponding to the different choices of $g$ overlap for all case studies considered. This demonstrates that the choice of $g$ (or $g_\beta$) has a negligible effect on our inference on $\eta$, not only for very large sample sizes, but even when the sample size is only moderately large. Somewhat surprisingly, this is also true when a parametric background proposal $g_\bt$, with $\bt$ estimated, is employed. This suggests that, even when the user has access to a testing procedure that can correctly account for the uncertainties associated with estimating  $g_\bt$ to achieve a reasonable approximation of $f_b$, the gain in power is negligible and the resulting inference is practically equivalent to an inferential analysis based on a potentially severely misspecified postulated background model $g$. This result not only provides strong evidence for the robustness of the proposed inferential strategy with respect to the choice of $g$, but also demonstrates that estimating the compensator suffices to account for any departure of $g$ (or $g_\beta$) from $f_b$, without affecting the resulting inference on $\eta$ and, therefore, minimizes its dependence on the scientist's guess of $g$.

\section{Inference without background-only data}
\label{sec:inf_wobkg}
Let us now consider the case in which a background-only sample is not available to the user. As we shall soon see, in this setup, the `closeness' of $g$ to $f_b$ plays a more prominent role. It is therefore sensible to focus on the parametric form of such a density, i.e., $g_{\bt}$. 

Since a background-only sample is not available, neither $\delta_\bt$ nor $\eta$ is estimable. Nevertheless, from \eqref{eq:eta_def}, we have that
\begin{equation}
    \label{eq:tht_0b_def}
    \tht_{0,\bt} = \frac{\tht_{\bt}}{\|S_\bt\|_{G_\bt}} \le \eta
\end{equation}
if $\delta_\bt \le 0$. In particular, inferring $\tht_{0,\bt}$ is equivalent to inferring $\eta$, when $\delta_\bt = 0$; whereas, testing and estimating $\tht_{0,\bt}$, in lieu of $\eta$, leads to a conservative, yet valid, inference whenever $\delta_\bt < 0$. Therefore, to avoid anti-conservative results, we can shift the focus of our inference to $\tht_{0,\bt}$ and identify suitable conditions under which $\delta_\bt\leq 0$. In what follows, we introduce one such condition for one-dimensional bump-hunting problems with smoothly decaying backgrounds, which are commonly encountered in physics and astronomy \citep[e.g.,][]{bumphunter, algeri2020testing,zhang_alageri_2023}. However, the proposed solution may fail to yield  conservative inference outside such a setting, where the background distribution may have an arbitrary shape.

\begin{figure}[t]
    \centering
    \includegraphics[width=0.48\linewidth]{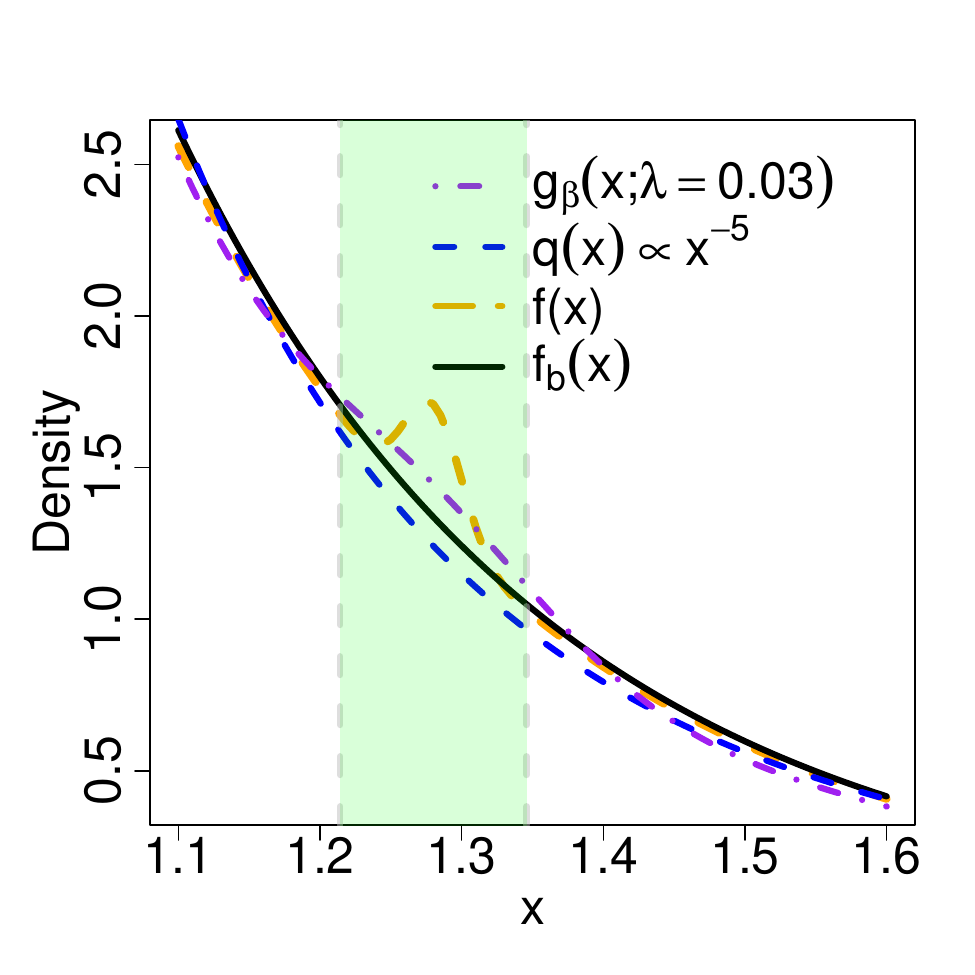}
    \includegraphics[width=0.48\linewidth]{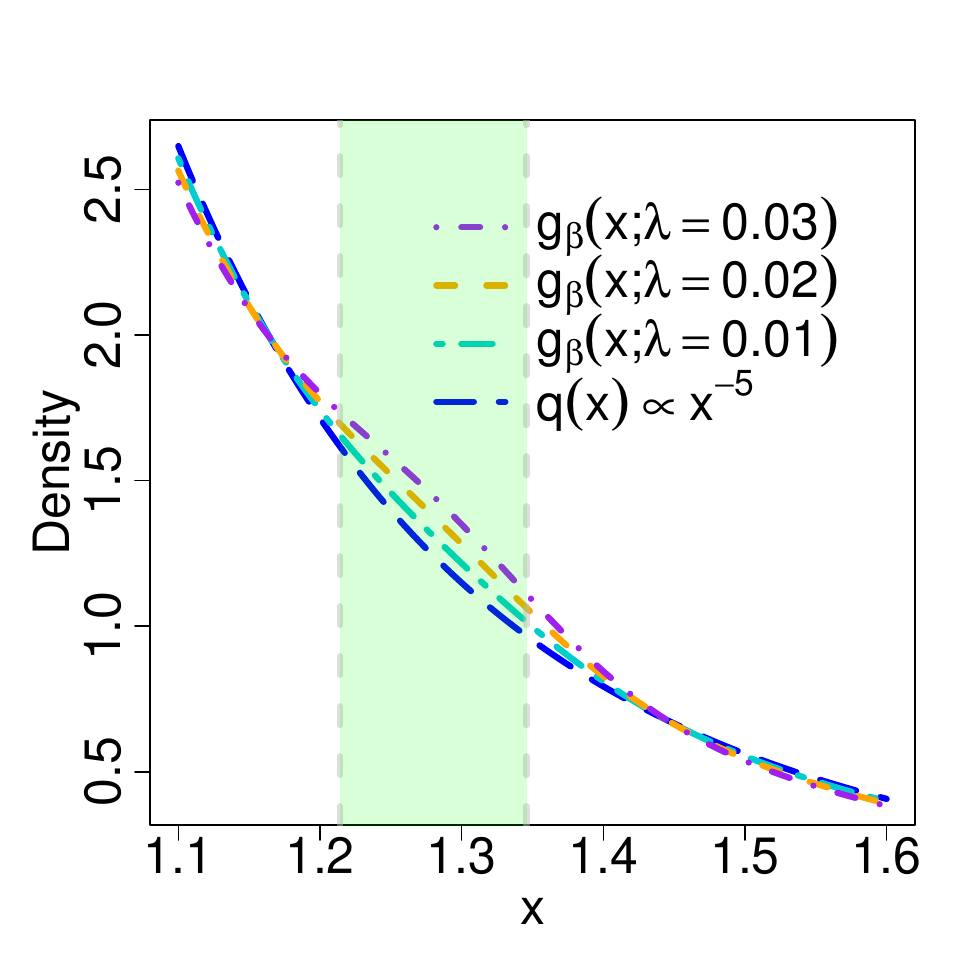}
  \caption{Left panel: Example of the data generating density $f$ (orange long-dashed line) for a bump-hunting scenario, the corresponding true background density $f_b$ (black solid line) as in \eqref{eq:LRT_sim_fb}, the baseline background model $q$ (blue dashed line) and the proposal background $g_\bt$ (purple dot-dashed line) with a bump added across the signal region (green shaded region). Right panel: Example of a sensitivity analysis with $g_\bt$ as in \eqref{eq:g_lambda} for different values of $\lambda$.
  }
  \label{fig:g_with_bump}
\end{figure}

Let the search region, $\mathcal{X}$, be a bounded interval on the real line.
In bump-hunting problems, the presence of the signal over such a region is characterized by a peak in the data-generating density $f$ (orange long-dashed line in Figure \ref{fig:g_with_bump}, left panel). In such cases, the signal distribution $F_s$ is unimodal, with mode $\mu_s$, has relatively low dispersion, and is typically modeled by a Gaussian or other bell-like symmetric distributions. As a result, $F_s$ places most of its mass over a `localized' signal region $M_\eps = [\mu_s - d_\eps, \mu_s + d_\eps]\subset \mathcal{X}$ (green shaded region in Figure \ref{fig:g_with_bump}) such that $F_s(M_\eps) = 1-\eps$ for some small $\eps>0$. Observe that
\[
    \delta_\bt \propto \intlu S_\bt(x)dF_b(x) = \Big\langle\frac{f_s}{g_\bt}, \frac{f_b}{g_\bt}\Big\rangle_{G_\bt} - 1,
\]
Therefore, $\delta_\bt\leq 0$ if the inner product in the above expression is at most one. Notice that:
\begin{equation}
    \label{eq:fs_g_fb_g_inner}
        \Big\langle\frac{f_s}{g_\bt}, \frac{f_b}{g_\bt}\Big\rangle_{G_\bt} = 
        \intlu \frac{f_s(x)}{g_\bt(x)} \frac{f_b(x)}{g_\bt(x)} dG_\bt(x) 
        \le 
        \sup_{x \notin M_\eps} \frac{f_b(x)}{g_\bt(x)} \eps + 
        \sup_{x \in M_\eps} \frac{f_b(x)}{g_\bt(x)} (1-\eps).
\end{equation}

For sufficiently small $\eps$, the first term on the right-hand side of the above inequality can be ignored for the moment (the reasoning behind this approximation is given below), whereas the structure of the second term implies that $\delta_\bt\leq 0$ whenever $\sup_{x \in M_\eps} \frac{f_b(x)}{g_\bt(x)} \le 1$. This last condition is satisfied whenever $g_\bt\geq f_b$ for all $x\in M_\eps$.

There are many possible ways to construct a density $g_\bt$ that satisfies such a condition in a bump-hunting scenario. The left panel of Figure \ref{fig:g_with_bump} illustrates one possible route based on the following heuristic: start with a baseline density, $q_\alpha$ (blue dashed line), possibly depending on some unknown parameter $\alpha\in \mathcal{A}\subset\R^{p-1}$, and let $g_\bt$ be a mixture inclusive of both $q_\alpha$ and a `dominating component', such as a diffused bump, enabling us to bound $f_b$ over $M_\epsilon$ (purple dot-dashed line). Note that, the baseline model $q_\al$ here, even though misspecified, is expected to provide a reasonable approximation of the smoothly decaying background so that: (i) the size of the dominating term being injected can indeed be calibrated to bound $f_b$, that is unknown, over $M_\epsilon$; (ii) having a $q_\alpha$ sufficiently close to the true background outside of $M_\eps$ ensures that $\sup_{x \notin M_\eps} \frac{f_b(x)}{g_\bt(x)}$ remains tractable. In such situations, it is reasonable to ignore the contribution from the first term on the right hand side of \eqref{eq:fs_g_fb_g_inner} when $\eps$ is sufficiently small (we recommend $\eps$ to be $O(10^{-3})$ in practice). Typical choices for such a baseline model $q_\alpha$ may involve a truncated exponential or a power-law density model (for reference see the Fermi-LAT example discussed in Section \ref{sec:data_analysis}). Additionally, the size of the dominating term cannot be excessively large; if it were, it could either lead to overly conservative inference or overly inflate $\sup_{x \notin M_\eps} \frac{f_b(x)}{g_\bt(x)}$, thereby failing to provide a negative $\del_\bt$ altogether.

In essence, $q_\alpha$ and the size of the dominating term should be such that $\delta_\bt$ approaches zero from below. Let $\bt= (\alpha, \lambda)$ in which $\alpha$ characterizes the shape of $q_\al$ and $\lambda$ controls the size of the dominating term introduced in $g_\bt$. Choose the latter so that:
\begin{equation}
    \label{eq:g_lambda}
    g_\bt(x) = (1-2\lambda) q_\al(x) + \lambda\Big[
    \phi_{\mathcal{X}}(x; \mu_1, \sigma_0) + \phi_{\mathcal{X}}(x; \mu_2, \sigma_0)
    \Big];
\end{equation}
with $\beta = (\al, \lambda) \in \mathcal{A} \times [0,1/2)$,  and $\phi_{\mathcal{X}}(\cdot; \mu, \sigma)$ denoting the Gaussian density with mean $\mu$ and standard deviation $\sigma$ truncated on $\mathcal{X}$. By placing two Gaussian densities centered at $\mu_1$ and $\mu_2$, respectively, both with scale parameter $\sigma_0$, the parametric form in \eqref{eq:g_lambda} allows us to inject a dominating component that visually appears as a diffused bump on top of $q_\alpha$. As shown through the numerical example and the real data analysis that follows, to ensure $f_b$ is dominated by \eqref{eq:g_lambda} over $M_\epsilon$ for at least some values of $\lambda$, it is recommended to choose $\mu_1$ and $\mu_2$ within $M_\eps$, but close to its boundaries $\mu_s - d_\eps$ and $\mu_s + d_\eps$, whereas $\sigma_0$ should be substantially larger than that of the width of the expected signal described by $f_s$. 
By varying $\lambda$, we can control the contribution of the dominating component.
Specifically, once $\al$ is estimated on the physics data via its MLE, $\widehat{\alpha}_n$, a sensitivity analysis for different values of $\lambda$ can then be performed to allow the user to visually assess, through plots such as the one presented on the right panel of Figure \ref{fig:g_with_bump}, which values of $\lambda$ are expected to yield dominance of $f_b$ over $M_\epsilon$. The corresponding values of the compensators are reported in Figure \ref{fig:power_curve_wobkg}.

Let $\lambda^*$ be the chosen value for $\lambda$ through the sensitivity analysis and let $\bts = (\als, \lambda^*)$, in which $\als$ denotes the minimizer of the KL divergence between the $q_\alpha$ and $f$. If  \ref{assump:gb_cont_x} - \ref{assump:bts_interior} hold for $\{q_\al\colon \al \in \mathcal{A} \}$, then $\alh_n \pconv \als$, and a conservative counterpart of the test in \eqref{eq:eta_test} is:

\begin{equation}
    \label{eq:tht_0bs_test}
    H_{0,\tht_0}: \tht_{0,\bts} \le 0 \quad \text{vs} \quad
    H_{1,\tht_0}: \tht_{0,\bts} >0.
\end{equation}

Let $\widehat{\theta}_{0,\widehat\beta_n}$ be the estimator of $\tht_{0,\bts}$ 
with $\widehat{\beta}_n=(\widehat{\alpha}_n, \lambda^*)$.  The test above can then be conducted on the basis of the following proposition.
\begin{prop}
\label{prop:thth_0b_dist}
    If \ref{assump:gb_cont_x} - \ref{assump:bts_interior} hold for $\{q_\al\colon \al \in \mathcal{A} \}$, then
    \begin{equation}
    \label{eq:prop_thth_0b_dist}
        \sqrt{n}(\thth_{0,\bth_n} - \tht_{0,\bts}) \dconv \mN\Big(0, \sig_{\bts, \tht_0}^2\Big), \quad \text{as $n \to \infty$,}
    \end{equation}
    where $\sig_{\bts, \tht_0}^2 = 
        \frac{\sig_{\bts, \tht}^2}{\|S_{\bts}\|_{G_{\bts}}^2} + 
        D_{\als}^T\mJ_{\als}^{-1}\mV_{\als}\mJ_{\als}^{-1}D_{\als} + 
        2D_{\als}^T \mJ_{\als}^{-1}\mC_{\als}$ and $
        \sig_{\bts,\tht}^2 = \intlu \big(\Sd_{\bts}(x)\big)^2dF(x) - \tht_{\bts}^2$, with $D_{\als}, \mC_{\als} \in \R^{p-1}$ and $\mJ_{\als}, \mV_{\als}$ are $(p-1) \times (p-1)$ matrices depending on $\als$. 
\end{prop}

The expressions for $D_{\als}, \mC_{\als}, \mJ_{\als}$ and $\mV_{\als}$ along with the proof of Proposition \ref{prop:thth_0b_dist} are given in Appendix \ref{app:prop_thth_0b_dist_proof}. $\mC_{\als}, \mJ_{\als}$, and $\mV_{\als}$ are analogous to $C_{\bts}$, $J_{\bts}$, and $V_{\bts}$ in Proposition \ref{prop:eta_unb_wbkg_beta_est}, except the former quantities involve $q_\al$ instead of $g_\bt$ and the expectations are taken under $F$ rather than $F_b$.

Given a consistent estimator $\sigh_{\bth_n, {\tht}_0}^2$  for $\sig_{\bts,\tht_0}^2$ (see Appendix \ref{app:sig_bts_tht0_est}), the test statistic corresponding to, asymptotically, the uniformly most powerful (UMP) test  for \eqref{eq:tht_0bs_test} is given by
\begin{equation}
\label{eq:tht_test_stat_unb_wobkg}
    \mZ_3 = \frac{\sqrt{n}\thth_{0,\bth_n}}{\sigh_{\bth_n, {\tht}_0}}
\end{equation}
and, from Proposition \ref{prop:thth_0b_dist} and Slutsky's theorem, $\mZ_3$ converges to the standard normal distribution under $H_{0,\tht_{0}}$, as $n \to \infty$.
The asymptotic $p$-value for the above test is then given by $1 - \Phi(\mZ_3^{obs})$, with $\mZ_3^{obs}$ being the observed value of $\mZ_3$ on the physics sample.

It is worth noting that, even though $\lambda^*$ is determined via the sensitivity analysis performed on the physics data, Proposition \ref{prop:thth_0b_dist} treats $\lambda^*$ as fixed. Here, we do not quantify the uncertainties arising from post-selection inference \citep{ps_inf}, as this is beyond the scope of this paper. As in most machine learning analyses, however, one could account for such uncertainties by splitting the data.

\begin{figure}[t]
    \includegraphics[width=0.48\textwidth]{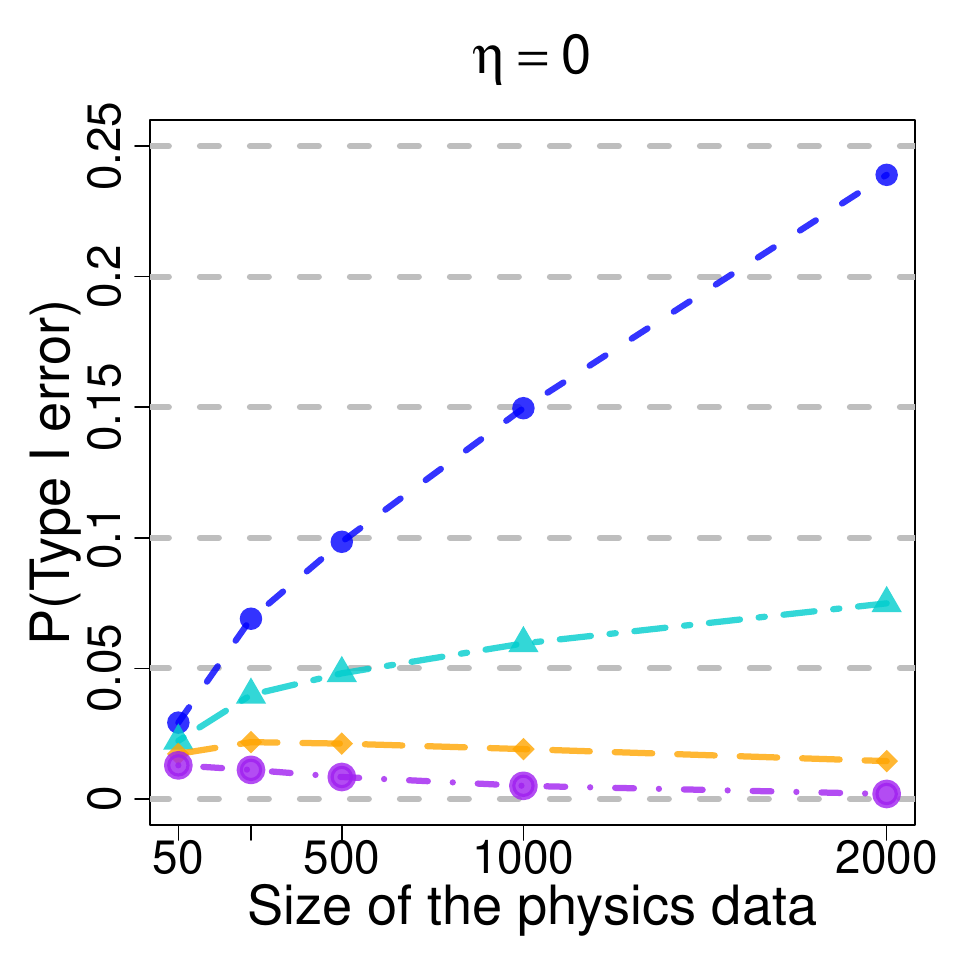}
       \includegraphics[width=0.48\textwidth]{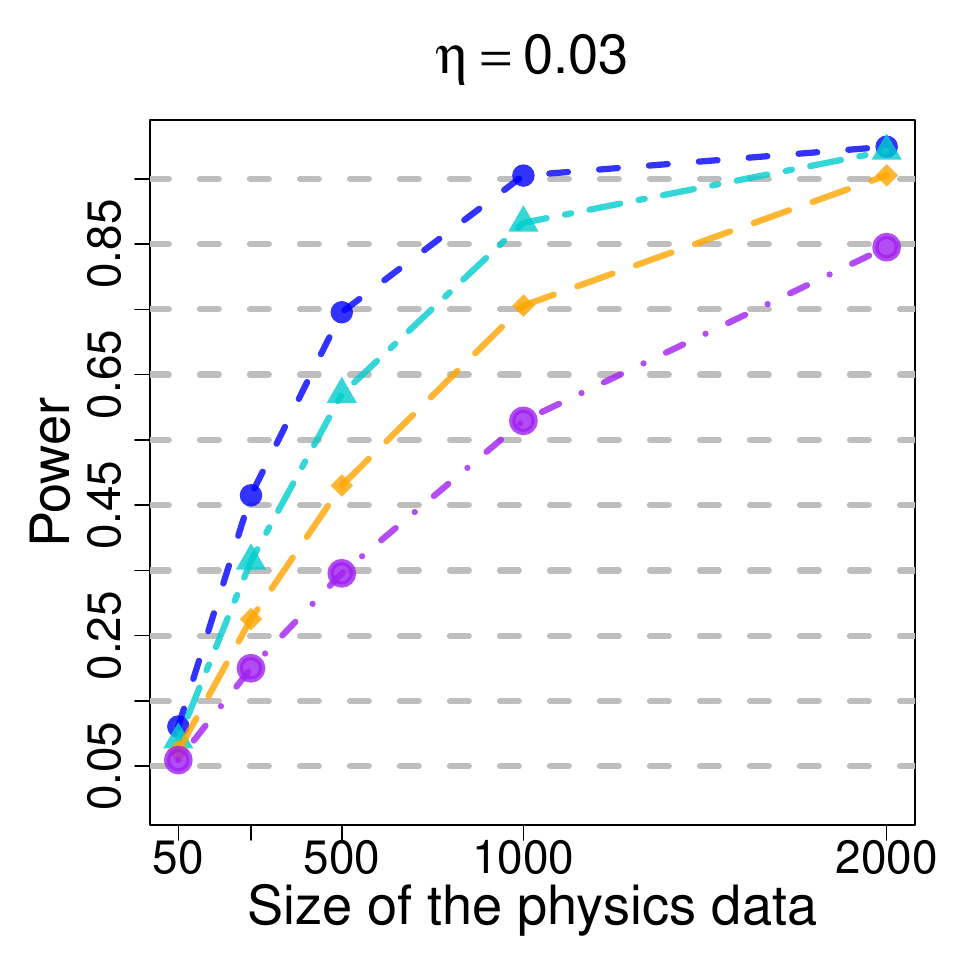}
\centering  \begin{subfigure}[b]{\textwidth}
    \centering
    \includegraphics[width=0.95\textwidth, height = 3cm]{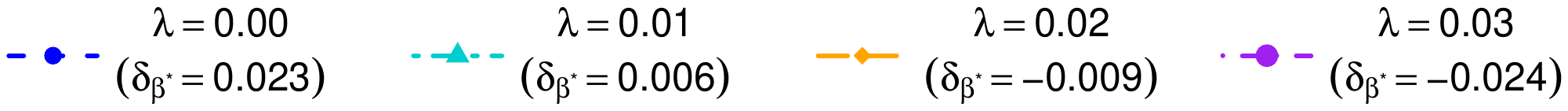}
  \end{subfigure}
  \caption{Simulated type I error probabilities (left panel) and power when $\eta = 0.03$ (right panel) for different choices of $\lambda$ and different sample sizes.} 
  \label{fig:power_curve_wobkg}
\end{figure}
\textbf{Numerical example.}
Let us now investigate the statistical properties of the test statistic in \eqref{eq:tht_test_stat_unb_wobkg} with an example. Choose $f_b$ and $f_s$ as in \eqref{eq:LRT_sim_fb} and \eqref{eq:LRT_sim_missp}, respectively. The type I error probability and the power at a 5\% significance level are obtained via a Monte Carlo simulation involving 100,000 replicates. For each such replicate, physics samples of size  $n=50, 250, 500, 1000$, and $2000$  are simulated from \eqref{eq:mix_model} for both $\eta = 0$ and $\eta =0.03$. The proposal background $g_\bt$ in \eqref{eq:g_lambda} is chosen so that $q_\al$ corresponds to a power-law density with shape parameter $\al$ estimated via MLE. Let $\eps = 0.001$ and set $\mu_1 = 1.25$, $\mu_2 = 1.31$, $\sig_0 = 0.08$. 
As shown in Figure \ref{fig:g_with_bump}, different values of $\lambda$ affect the prominence of the dominating component; hence, the need to assess the performance of the test in \eqref{eq:tht_test_stat_unb_wobkg} for each of such values.

Figure \ref{fig:power_curve_wobkg} shows the simulated type I error probabilities (left panel) and the simulated power for $\eta=0.03$ (right panel). Additional power curves for $\eta = 0.01$ and $\eta = 0.02$ are given in Appendix \ref{app:additional figures}.  For $\lambda = 0$ and $\lambda = 0.01$, the compensator is positive, indicating anti-conservativeness of the corresponding tests. This is confirmed by the left panel of Figure \ref{fig:power_curve_wobkg} in which the corresponding type I error probabilities exceed 0.05 (blue dashed line and cyan double dashed line). Conversely, for $\lambda = 0.02$ (orange long-dashed line) and $\lambda = 0.03$ (purple dot-dashed line), corresponding to the case in which $g_\bt$ exhibits a more prominent dominating component, the compensator is negative. As expected, in these last two cases, the resulting tests are conservative, and the corresponding type I error probabilities remain well below the nominal threshold of 0.05. 

Note that, for all cases considered, the type I error probabilities do not approach the nominal level of 0.05 as the sample size increases. Instead, they diverge for $\delta_{\bts}>0$ and approach zero for $\delta_{\bts}<0$. This is a natural consequence of Proposition \ref{prop:thth_0b_dist}. In particular, the type I error probability:
\[
    1-\Phi\Big(z_{0.05} - \frac{\del_{\bts}\sqrt{n}}{\sig_{\bts, \tht}}\Big),
\]
with $z_{0.05}$ denoting the upper $5\%$ point of the standard normal distribution, is an increasing function of $n$ when $\delta_{\bts}>0$ and it is a decreasing function of $n$ for $\delta_{\bts}<0$. This, however, does not imply that the procedure is more conservative as the sample size increases: the right panel of Figure \ref{fig:power_curve_wobkg} demonstrates that, regardless of the sign of $\delta_{\bts}$, the power increases with the sample size. Moreover, as expected, as $\lambda$ increases, the compensator becomes increasingly negative and the corresponding tests become more conservative, leading to a drop in the power for any fixed sample size as $\lambda$ varies from $0$ to $0.03$.

\section{Analysis on Fermi-LAT Data}
\label{sec:data_analysis}
In this section, we illustrate the implementation of the tests in \eqref{eq:eta_test_stat_unb_wbkg_b_knw}, \eqref{eq:eta_test_stat_unb_wbkg_b_est}, and \eqref{eq:tht_test_stat_unb_wobkg} using a realistic simulated dataset from the Fermi Large Area Telescope \citep{atwood2009large}, under both the settings in which a background-only sample is either accessible or inaccessible to the user.
The data considered, previously analyzed in \citet{algeri2021testing}, mimic a mixture of $\gamma$-ray signals produced by both dark matter annihilation and astrophysical background.
The background distribution in this scenario is a power-law distribution indexed by the parameter $\psi$, whereas the signal events are generated from a Gaussian distribution with mean $\kappa$ and  standard deviation $\kappa/10$. Additional perturbations due to systematic effects and instrumental errors of nature unknown to us are also present in the data. The search region consists of the energy interval $[1,35]$ gigaelectronvolts (GeV). 

In the proposed analysis, the data is transformed into the log-scale to ease visualization in the sensitivity analysis that follows. The data-generating density for the transformed data is then given by $f(x; \eta)$ in \eqref{eq:mix_model} with
\begin{equation}
\label{eq:fermi_lat_setup}
   f_b(x) \propto \exp(-x\psi)  \quad \text{and}\quad f_s(x) \propto \exp\Big\{-\frac{(\exp(x)-\kappa)^2}{0.02\kappa^2}\Big\} \exp(x),    
\end{equation}
with $x\in [0,\log(35)]$.
The physics and background-only samples have sizes $n=2338$ and $m=4427$, respectively, and were generated by setting $\kappa = 3.5$ and $\psi = 1.4$.
\begin{figure}[t]
    \centering
    \includegraphics[width=0.48\linewidth]{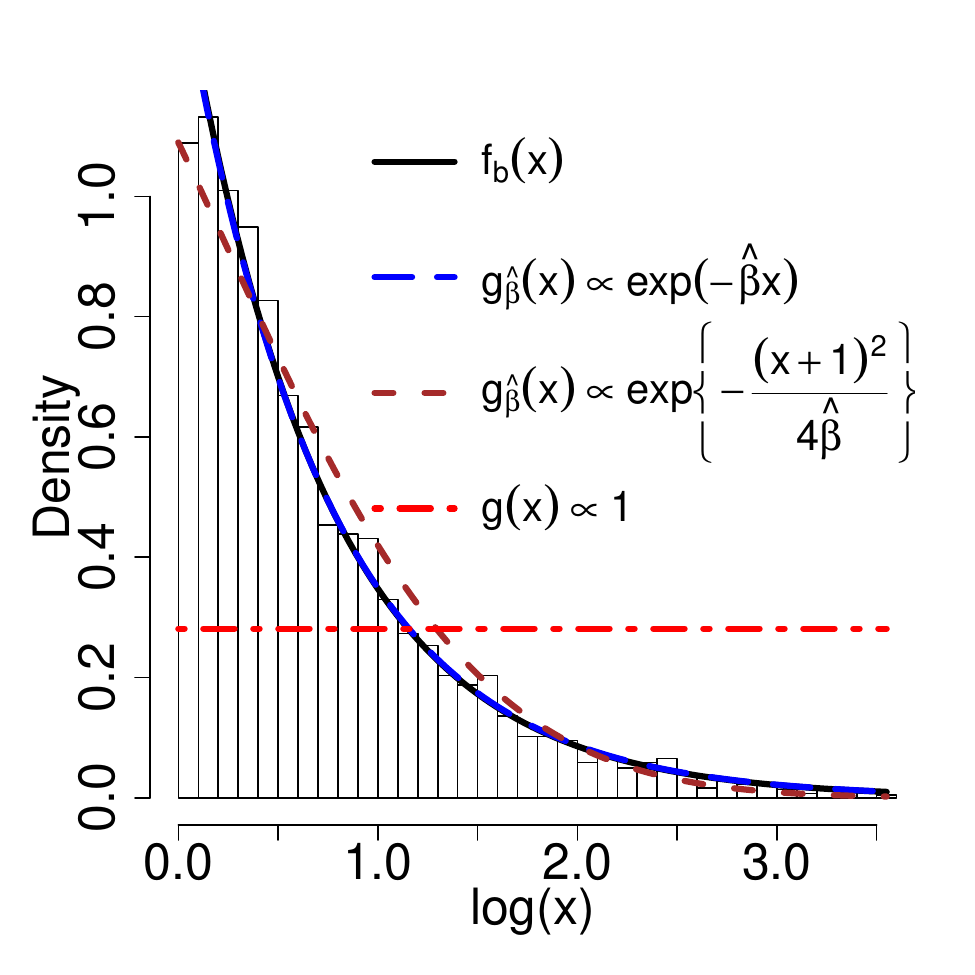}
    \includegraphics[width=0.48\linewidth]{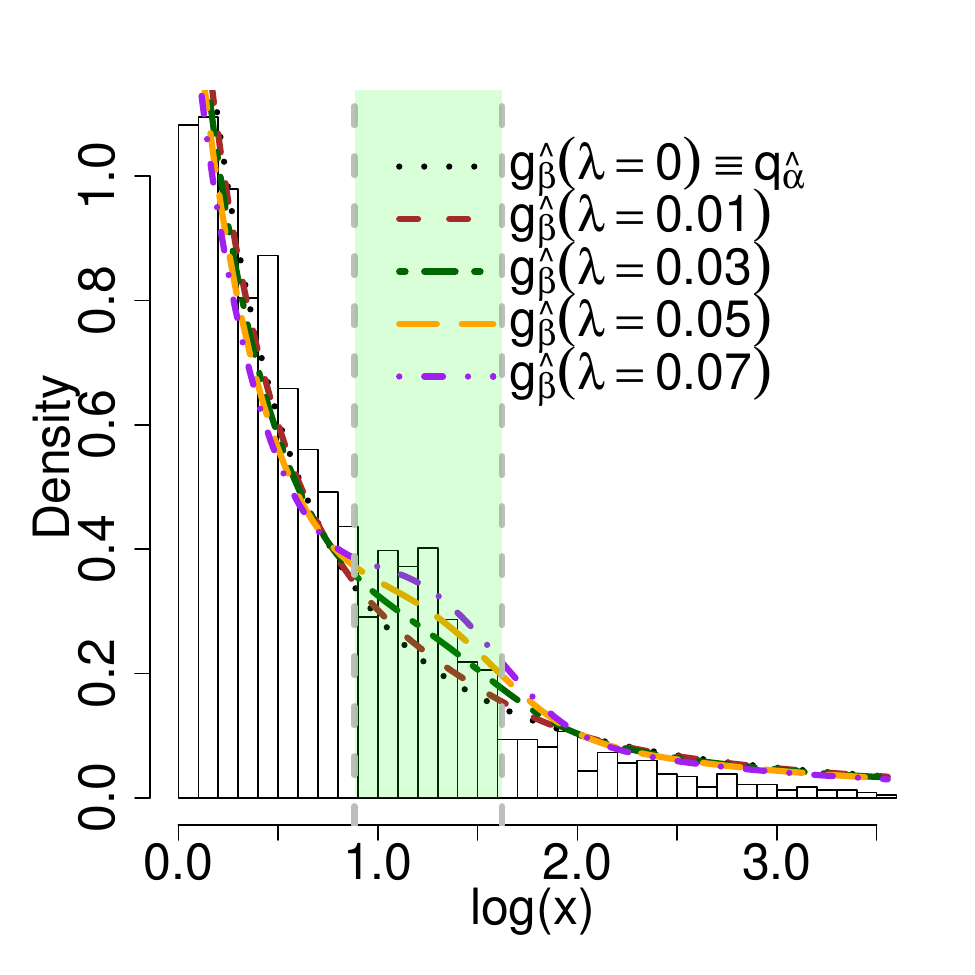}
    \caption{Left panel: Graph of a uniform distribution (red double dashed line) as $g$ along with an exponential distribution (blue long-dashed line) and a Gaussian tail (brown dashed line) as parametric proposal backgrounds $g_\bt$, $\bt$ estimated via MLE, overlaid on the histogram of the background-only sample in log-scale. Right panel: Sensitivity analysis with different levels of $\lambda$ in $g_{\bth_n}$, overlaid on the histogram of the physics data in log-scale.}
    \label{fig:Fermi_LAT_g}
\end{figure} 
The signal search in the presence of a background-only sample is then performed with three different choices of the proposal background. Those correspond to a uniform distribution over the search region $[0, \log(35)]$ and two parametric forms for $g_\bt$: the true background $f_b$ in \eqref{eq:fermi_lat_setup}, with $\bt=\psi$ treated as unknown, and 
\begin{equation*}
\label{eq:g_bt_gauss_tail}
    g_\bt(x) \propto \exp\Big\{-\frac{(x+1)^2}{4\bt}\Big\}, \quad x \in [0, \log(35)],
\end{equation*}
corresponding to the tail of a Gaussian density with width controlled by $\bt$. In both cases, the MLE of $\bt$ is calculated on the background-only sample. As shown in Figure \ref{fig:Fermi_LAT_g}, the above choices for the proposal background distributions exhibit different degrees of deviation from $f_b$ (black solid line). The exponential background (blue long-dashed line), having the same parametric form as $f_b$, provides the closest approximation.
\begin{table}[t]
    \centering
 \begin{tabular}{|c|c|c|c|}
\hline
     \textbf{Proposal background} & $\mathbf{\widehat\eta}$ & 95\% \textbf{Confidence Interval} & $p$\textbf{-value}\\ \hline
     Exponential & 0.0426 & (0.0252, 0.0601) & $8.524 \times 10^{-7}$\\
     \hline
     Gaussian-tail & 0.0426 & (0.0251, 0.0600) & $8.625 \times 10^{-7}$\\
     \hline
     Uniform & 0.0433 & (0.0257, 0.0608) & $6.622 \times 10^{-7}$\\
     \hline
\end{tabular}
    \caption{Comparison of signal search results on the Fermi-LAT data in the presence of a background-only data for different choices of the proposal background.}
    \label{tab:fermi_lat_wbkg}
\end{table}

The inferential results of the signal search conducted as described in Section \ref{sec:inf_unbinned_wbkg} for each choice of the proposal background are summarized in Table \ref{tab:fermi_lat_wbkg}. Both parametric choices for $g_\bt$ yield $\hat\eta = 0.0426$, and the uniform proposal background leads to $\hat\eta = 0.0433$, with all having  $p$-values of order $O(10^{-7})$, demonstrating, once again, the robustness of the proposed approach with respect to the choice of $g$ (or $g_\beta$). The 95\% confidence intervals reported in Table \ref{tab:fermi_lat_wbkg} across all three choices are also similar.  We note that the results obtained are also comparable to those obtained in \citet{algeri2021testing} ($\widehat{\eta}=0.045$, $p$-value=$2.11\cdot 10^{-8}$) in which the same data is analyzed with a likelihood-based approach that assumes $f_b$ in  \eqref{eq:fermi_lat_setup} is known. 

Recall that, although the signal search presented here is one-dimensional, the proposed methodology can also be applied to multidimensional settings. The one-dimensional example helps us compare the signal search results presented in Table \ref{tab:fermi_lat_wbkg} with the following analysis in the absence of background-only data.

Now we assume only the physics sample is available, and proceed with the sensitivity analysis described in Section \ref{sec:inf_wobkg}. We use the parametric form for $g_\bt$ described in \eqref{eq:g_lambda} with a shifted power-law density given by:
\begin{equation}
\label{eq:q_al_Fermi_LAT}
    q_\al(x) \propto (x+1)^{-(\al+1)},\quad x\in[0 ,\log(35)], 
\end{equation}
as our baseline background model. Let $\eps = 0.001$, $\mu_1 = 1.07$, $\mu_2 = 1.44$, and $\sig_0 = 0.31$. 
 The MLE of $\al$, obtained on the physics data, is $\alh_n = 1.59$. A sensitivity analysis can then be performed on the basis of the right panel of Figure \ref{fig:Fermi_LAT_g}. Diffuse bumps are added on top of $q_{\alh_n}$ (black dotted line), across the signal region (green shaded region) following the approach described in Section \ref{sec:inf_wobkg} and setting $\lambda = 0.01$ (brown dashed line), $\lambda = 0.03$ (green double dashed line), $\lambda = 0.05$ (orange long-dashed line) and $\lambda = 0.07$ (purple dot-dashed line).

Observe that the dominating component in $g_\bt$ becomes visible at $\lambda = 0.03$ and a noticeable change of concavity occurs. Therefore, $g_{\bth_n}$ with $\bth_n = (\alh_n, 0.03)$ can reasonably be assumed to dominate $f_b$ in $M_\eps$, leading to a negative value for $\del_{\bts}$. Performing the test in \eqref{eq:tht_0bs_test} with $\lambda = 0.03$ leads to the conservative estimate $\thth_{0,\bth_n} = 0.0359$ with a $p$-value of $7.391 \times 10^{-7}$, similar to those reported in Table \ref{tab:fermi_lat_wbkg}. One could also consider the case $\lambda = 0.05$ with a more pronounced dominating component injected into $g_\bt$, ensuring a negative $\del_{\bts}$. The corresponding estimate, $\thth_{0,\bth_n} = 0.0232$, is, unsurprisingly, substantially lower with a larger $p$-value of order $O(10^{-3})$. Here, we still claim a discovery, but with a more conservative estimate and lower significance. For $\lambda = 0.07$, visual inspection suggests that the dominating component may be too large to enable any signal detection and, indeed,  the corresponding test leads to $\thth_{0,\bth_n} = 0.0100$ with a $p$-value of 0.0954. 

\section{Summary and extensions}
\label{sec:discussion}
The main contribution of this article is to unveil the existence of the compensator $\delta$: a single parameter that, when inferring the true signal intensity, collects all the necessary information needed to account for the discrepancy between the true background distribution and the one postulated by scientists. 

When a (labeled) background-only sample is available to the practitioners, the compensator is estimable, allowing for accurate inference on the signal intensity that is robust to the choice of the postulated background. 

The proposed method does not directly apply when the functional form of $F_s$ is unknown. In this setting, model-independent procedures \citep[e.g.,][]{purvasha_chakr} are recommended. 
Note that the test proposed in Section \ref{sec:inf_unbinned_wbkg} also assumes the background-only sample is well specified, i.e., generated from the true background $f_b$. 
When the background-only sample is believed to be misspecified or is unavailable to the users, the compensator is no longer estimable. In such situations, however, at least for the one-dimensional bump-hunting problems, the compensator enables us to establish a sufficient condition on the choice of the postulated background, $g_\beta$, that guarantees conservative, yet valid inference. Section \ref{sec:inf_wobkg} describes how to perform a sensitivity analysis to evaluate the validity of such a condition for different choices of $g_\beta$ and assess the degree of conservativeness of the resulting inference.  In the multidimensional setting, such a sensitivity analysis can still be performed `one-dimension-at-a-time', but extensions to the case in which $F_s$ depends on the unknown parameter $\gamma$ may require a more delicate treatment as the choice of $g_\beta$ may vary for different values of $\gamma$. 

\section*{Supplementary material and data availability}
Technical proofs and additional figures are available in the Appendix. The \texttt{R} code and the data that support the findings of this manuscript are openly available at \href{https://github.com/baner175/signal_detection}{https://github.com/baner175/signal\_detection}.

\section*{Acknowledgments}
The authors are grateful to Oliver Rieger and Lydia Brenner for the valuable feedback and discussions. These interactions motivated the thorough investigation of the background-only case described in Section \ref{sec:inf_unbinned_wbkg}.  

AB's work has been partially funded by the Errett W. McDiarmid Graduate Fellowship, provided by the College of Liberal Arts at the University of Minnesota. 
SA's work has been partially funded by the Warwick MidCareer Faculty Research Award, College of Liberal Arts, University of Minnesota.

\appendix

\section{Technical proofs}
\label{app:appendix}

\subsection{Proof for completeness of $\mT$}
\label{app:completeness}
Let $\mT^* = \{1, T^*_1, T^*_2, \cdots\}$ be a complete  orthonormal basis for $\Hl$. To prove the completeness of $\mT= \{1, \Sd, T_1, T_2, \cdots\}$, recall the following result:

\begin{result}\citep[p.~16]{conway2019course} An orthonormal set of functions $\mU=\{U_\ell\}_{\ell\geq 0}$ forms a complete basis for $\Hl$ if and only if, for some $h \in \Hl$, $\langle u_\ell , h\rangle_G=0$, for all $\ell\geq 0$, $\implies$ $h = 0$.  
\end{result}

Let $h \in \Hl$ be such that: 
\begin{equation}
    \label{eq:u_perp_mT}
    \langle h, 1\rangle_G = 0; \quad \langle h, \Sd\rangle_G = 0, \quad \text{and} \quad \langle h, T_j\rangle_G = 0\quad\text{for all  $j \ge 1$}.
\end{equation}

By construction, the functions $T_j\in \mT$, $j\geq 1$, specify as:
\begin{equation}
\label{eq:GS_orth}
    \begin{aligned}
        T_1^\circ =& \Ts_1 - \langle \Ts_1, \Sd\rangle_{G} \Sd; \quad T_1 = \frac{T_1^\circ}{\|T_1^\circ\|_G}\\
        \Tc_j =& \Ts_j - \sum_{k=1}^{j-1} \langle T_k, \Ts_j\rangle_G T_k - \langle \Sd, \Ts_j\rangle_G \Sd; 
    \quad T_j = \frac{T_j^\circ}{\|T_j^\circ\|_G}
    \end{aligned}
\end{equation}
Therefore, from \eqref{eq:u_perp_mT}-\eqref{eq:GS_orth}, it  follows that, $\langle h, 1 \rangle_G = 0=\langle h, \Ts_j \rangle_G $, for all $j \ge 1$ and since $\mT^*$ is complete and orthonormal,  $h = 0$, which completes the proof.
\hfill $\square$

\subsection{Proof of Proposition \ref{prop:eta_unb_wbkg}}
\label{app:prop_eta_unb_wbkg_proof}
Consider $\nu = (\tht_0, \del_0)$  in which, letting $S_0(x) = \frac{\Sd(x)}{\normS} = \frac{S(x)}{\normS^2}$,
\begin{equation}
    \label{eq:tht0_del0_def}
\theta_0 = \frac{\theta}{\normS} = \intlu S_0(x)dF(x); \quad \delta_0 = \frac{\delta}{\normS} = \intlu S_0(x)dF_b(x).
\end{equation}
The empirical estimator of $\nu$ is $\widehat\nu = (\thth_0, \delh_0)$ with
\begin{equation}
\label{eq:thth0_delh0_def}
    \widehat\theta_0 = \frac{\widehat\theta}{\normS} = \frac{1}{n}\sum_{i=1}^n S_0(X_i) \quad \text{and} \quad 
\widehat\delta_0 = \frac{\widehat\delta}{\normS} = \frac{1}{m}\sum_{i=1}^m S_0(Y_i).
\end{equation}

Under \eqref{eq:m_n_lim}, by the law of large numbers, we have $\widehat\nu \pconv \nu$,
 and by the central limit theorem:
\begin{equation}
    \label{eq:tht0_del0_dist_conv}
    \sqrt{n}(\thth_0 - \tht_0) \stackrel{d}\longrightarrow \mN \left(0, \sig^2_{\tht_0}\right)  \quad\text{and}\quad
    \sqrt{m}(\delh_0 - \del_0) \stackrel{d}\longrightarrow \mN \left(0, \sig^2_{\del_0}\right),
\end{equation}
where $\sig_{\tht_0}^2 = \frac{\sig_{\tht}^2}{\normS^2}$ and 
$\sig_{\del_0}^2 = \frac{\sig_{\del}^2}{\normS^2}$. The limiting distributions in \eqref{eq:tht0_del0_dist_conv} are valid if the second-order moments of $\thth_0$ and $\delh_0$ exist. This can be easily shown as the function $S_0$---being continuous on a compact set---is bounded.
Let us now consider the function
\begin{equation}
    \label{eq:T_def}
    T(x, y) = \frac{x-y}{1-y}.
\end{equation}
such that $\eta = T(\nu)$ and $\widehat\eta = T(\widehat\nu)$. We denote the gradient vector of $T$ by $\partial T$, with elements $\partial_i T$, for $i=1,2$, i.e., 
\begin{equation}
\label{eq:pT_def}
\begin{aligned}
    \partial_1T(x,y) = \frac{1}{1-y}\quad \text{and}\quad
    \partial_2T(x,y) = \frac{x-1}{(1-y)^2}.
\end{aligned}
\end{equation}
The mean value theorem gives:
\[
\begin{aligned}
    \widehat\eta - \eta =& T(\hat\nu) - T(\nu) = (\widehat\nu - \nu)^T\partial T(\tilde\nu) \\
    \implies 
    \sqrt{\frac{mn}{m+n}}\left(\widehat\eta - \eta\right) =&
    \sqrt{n}(\thth_0 - \tht_0) \sqrt{\frac{m}{m+n}}\partial_1 T(\tilde\nu) +
    \sqrt{m}(\delh_0 - \del_0) \sqrt{\frac{n}{m+n}}\partial_2 T(\tilde\nu)
\end{aligned}
\]
where $\tilde\nu$ is a point on the line segment joining $\nu$ and $\widehat\nu$. Since $T$ is continuously differentiable and $\widehat\nu$ is consistent for $\nu$, then $\partial T(\tilde\nu) \pconv \partial T(\nu)$. Moreover, given that $\thth_0$ and $\delh_0$ are independent, from \eqref{eq:tht0_del0_dist_conv} and Slutsky's theorem we obtain \eqref{eq:prop_eta_unb_wbkg_conv} with 
\[
\begin{aligned}
    \sig_\eta^2 =& (1-\pi) \partial_1 T(\tilde\nu)^2\sig_{\tht_0}^2 + \pi \partial_2T(\tilde\nu)^2\sig_{\del_0}^2\\
    =&
    \frac{(1-\pi)\sig_{\tht_0}^2}{(1-\del_0)^2} + \frac{\pi\sig_{\del_0}^2(\tht_0-1)^2}{(1-\del_0)^4}
    =
    \frac{(1-\pi)\sig_{\tht}^2}{(\normS-\delta)^2} + \frac{\pi(\tht - \normS)^2\sig_{\del}^2}{(\normS - \del)^4}.
\end{aligned}
\]
\hfill $\square$

\subsection{Proof of Proposition \ref{prop:eta_unb_wbkg_beta_est}}
\label{app:prop_eta_unb_wbkg_beta_est_proof}
Consider $\nu_\bt = (\tht_{0,\bt}, \del_{0,\bt})$ in which 
\begin{equation}
    \label{eq:tht0_b_del0_b_def}
    \tht_{0,\bt} = \frac{\tht_{\bt}}{\|S_\bt\|_{G_\bt}} = \intlu S_{0,\bt}(x)dF(x); \quad 
    \del_{0,\bt} = \frac{\del_{\bt}}{\|S_\bt\|_{G_\bt}} = \intlu S_{0,\bt}(x)dF_b(x)
\end{equation}
and 
\begin{equation}
\label{eq:S_0b_def}
    S_{0,\bt}(x) = \frac{\Sd_\bt (x)}{\|S_\bt\|_{G_\bt}} = \frac{S_\bt (x)}{\|S_\bt\|^2_{G_\bt}}
\end{equation}

for any $\bt \in \mB$.  Let $\widehat\nu_\bt = (\thth_{0,\bt}, \delh_{0,\bt})$ be the estimator of $\nu_\bt$ with elements:
\begin{equation}
    \label{eq:thth0_b_delh0_b_def}
    \thth_{0,\bt} = \frac{\thth_{\bt}}{\|S_\bt\|_{G_\bt}} = \frac{1}{n}\sum_{i=1}^n S_{0,\bt}(X_i), \quad 
    \delh_{0,\bt} = \frac{\delh_{\bt}}{\|S_\bt\|_{G_\bt}} = \frac{1}{m}\sum_{i=1}^n S_{0,\bt}(Y_i),
\end{equation}
and such that, for any fixed $\bt$, $\widehat\nu_\bt \pconv \nu_\bt$.

Recall that  $\bts$ denotes the minimizer of the KL divergence between $G_\beta$ and $F_b$. Then, $\eta = T(\nu_{\bts})$ and $\widehat\eta_{\bth_m} = T(\widehat\nu_{\bth_m})$ with $T$ as in \eqref{eq:T_def}. In what follows, $\partial_\bt$ and $\partial^2_\bt$ denote, respectively, the $p$-dimensional gradient vector and the $p \times p$ Hessian matrix of a function with respect to $\bt$. 
Assumptions \ref{assump:gb_cont_bt}-\ref{assump:bts_interior} ensure that, under $F_b$, $\bth_m$ is a consistent estimator for  $\bts$  and allow us to obtain the following expansion:
\begin{equation}
    \label{eq:bthm_taylor}
    \sqrt{m}(\bth_m - \bts) = 
    J_{\bts}^{-1}
    \frac{1}{\sqrt{m}}\left(\summ \Big[\pb\log g_{\bt}(Y_i)\Big]_{\bt = \tilde\bt}\right) + o_P(1)
\end{equation}
where $J_{\bts} = \mE_{F_b}\left[-\ppb\log g_{\bt}(Y)\right]\big|_{\bt = \bts}$. Next, we apply the mean value theorem on $\thth_{0,\bt}$ at $\bth_m$ about $\bts$ to get:
\begin{equation}
    \label{eq:thth0b_taylor}
    \begin{aligned}
        \thth_{0,\bth_m} &= \thth_{0,\bts} + 
        \left[\pb\thth_{0,\bt}\right]_{\bt = \tilde\bt}^T (\bth_m - \bts)\\ \implies 
        \thth_{0,\bth_m} - \tht_{0,\bts} &= 
        (\thth_{0,\bts} - \tht_{0,\bts}) + 
        \left[\pb\thth_{0,\bt}\right]_{\bt = \tilde\bt}^T (\bth_m - \bts)
    \end{aligned}
\end{equation}
where $\tilde\bt$ is a point lying on the line joining $\bth_m$ and $\bts$, hence, $\tilde\bt \pconv \bts$. Assumptions \ref{assump:gb_cont_x}-\ref{assump:bts_interior} guarantee uniform probability convergence of the relevant functions (see \citealt[p.~46]{Vaart_1998}), thereby allowing us to replace the sample average with the corresponding expectation and $\tilde\bt$ or $\bth_m$ with $\bts$ in the asymptotic limit simultaneously. Moreover, for any fixed $\bt$, by \ref{assump:gb_cont_x}, $\pb S_{0,\bt}(x)$ is bounded, hence, by the dominated convergence theorem, we can interchange the differentiation in $\bt$ and integral with respect to $F$. Thus,
\begin{equation}
    \label{eq:pb_thth0b_pconv}
    \pb\thth_{0,\tilde\bt} = \frac{1}{n}\sumn \Big[\pb S_{0,\bt}(X_i)\Big]_{\bt = \tilde\bt}\pconv \intlu \Big[\pb S_{0,\bt}(x)\Big]_{\bt = \bts}dF(x) = \pb \tht_{0,\bts}.
\end{equation}
Combining \eqref{eq:thth0_b_delh0_b_def}-\eqref{eq:pb_thth0b_pconv} gives:
\begin{equation}
    \label{eq:thth0b_expansion}
\begin{aligned}
    \sqrt{\frac{mn}{m+n}}\left(\thth_{0,\bth_m} -\tht_{0,\bts}\right) =&
    \sqrt{\frac{m}{m+n}}\frac{1}{\sqrt{n}}\sumn\left(S_{0,\bts}(X_i) - \tht_{0,\bts}\right)+ \\ &
    \quad \sqrt{\frac{n}{m+n}}\Big[\pb\tht_{0,\bt}\Big]_{\bt = \bts}^T J_{\bts}^{-1} \frac{1}{\sqrt{m}}\summ\Big[\pb\log g_{\bt}(Y_i)\Big]_{\bt = \bts} + o_P(1).
\end{aligned}
\end{equation}

Similarly, for $\delh_{0,\bth_m}$ we have:
\begin{equation}
    \label{eq:delh0b_expansion}
\begin{aligned}
    \sqrt{\frac{mn}{m+n}}\left(\delh_{0,\bth_m} -\del_{0,\bts}\right) =&
    \sqrt{\frac{n}{m+n}}\frac{1}{\sqrt{m}}\sumn\left(S_{0,\bts}(Y_i) - \del_{0,\bts}\right)+ \\ &
    \quad \sqrt{\frac{n}{m+n}}\Big[\pb\del_{0,\bt}\Big]_{\bt = \bts}^T J_{\bts}^{-1} \frac{1}{\sqrt{m}}\summ\pb\Big[\log g_{\bt}(Y_i)\Big]_{\bt = \bts} + o_P(1).
\end{aligned}
\end{equation}
Applying the mean value theorem to  $T$ gives:
\begin{equation}
    \label{eq:eth_bth_taylor}
    \begin{aligned}
    \sqrt{\frac{mn}{m+n}}(\widehat\eta_{\bth_m} - \eta) =& \sqrt{\frac{mn}{m+n}}\left(T(\widehat\nu_{\bth_m}) - T(\nu_{\bts})\right)\\
    =&
    \sqrt{\frac{mn}{m+n}}\pone T(\nut)\left(\thth_{0,\bth_m} -\tht_{0,\bts}\right) + 
    \sqrt{\frac{mn}{m+n}}\ptwo T(\nut)\left(\delh_{0,\bth_m} -\del_{0,\bts}\right) \\
\end{aligned}
\end{equation}
with $\pone T$ and $\ptwo T$ defined as in \eqref{eq:pT_def} and $\nut$ is a point lying on the line segment joining $\widehat\nu_{\bth_m}$ and $\nu_{\bts}$. A similar expansion to \eqref{eq:thth0b_taylor} holds for $\delh_{0,\bth_m}$. Hence, $\thth_{0,\bth_m} \pconv \tht_{0,\bts}$, $\delh_{0,\bth_m} \pconv \del_{0,\bts}$, $\widehat\nu_{\bth_m} \pconv \nu_{\bts}$, and consequently $\nut \pconv \nu_{\bts}$. Combining \eqref{eq:eth_bth_taylor} with \eqref{eq:thth0b_expansion}-\eqref{eq:delh0b_expansion} gives:
\begin{equation}
    \label{eq:etahb_expansion}
    \begin{aligned}
    \sqrt{\frac{mn}{m+n}}(\widehat\eta_{\bth_m} - \eta) &=
    \sqrt{\frac{m}{m+n}}\pone T(\nut)\left[\frac{1}{\sqrt{n}}\sumn\left(S_{0,\bts}(X_i) - \tht_{0,\bts}\right) \right]+ \\
    &\quad 
    \sqrt{\frac{n}{m+n}}\ptwo T(\nut)\left[\frac{1}{\sqrt{m}}\summ\left(S_{0,\bts}(Y_i) - \tht_{0,\bts}\right)\right] +\\
    & \quad
    \sqrt{\frac{n}{m+n}}\Gamma_{\bts}^TJ_{\bts}^{-1}
    \left[\frac{1}{\sqrt{m}}\summ\pb\log g_{\bt}(Y_i)\right]_{\bt = \bts}
    + o_P(1)
\end{aligned}
\end{equation}
with
\[
\begin{aligned}
    \Gamma_{\bts} =& 
    \pone T(\nu_{\bts}) \Big[\pb\tht_{0,\bt}\Big]_{\bt = \bts} + 
    \ptwo T(\nu_{\bts}) \Big[\pb\del_{0,\bt}\Big]_{\bt = \bts} \\ 
    =& 
    \left[\frac{1}{1-\del_{0,\bts}} \Big[\pb\tht_{0,\bt}\Big]_{\bt = \bts} + 
    \frac{(\tht_{0,\bts} - 1)}{(1-\del_{0,\bts})^2} \Big[\pb\del_{0,\bt}\Big]_{\bt = \bts}
    \right].
\end{aligned}
\]

Let us now establish the asymptotic distributions of each of the summands on the right-hand side of \eqref{eq:etahb_expansion}. By the central limit theorem and Slutsky's theorem, the asymptotic distribution of the first summand is:
\begin{equation}
\label{eq:eta_hat_b_dist_1}
    \sqrt{\frac{m}{m+n}} \pone T(\nut)
    \left[\frac{1}{\sqrt{n}}\sumn\left(S_{0,\bts}(X_i) - \tht_{0,\bts}\right)\right]
    \dconv \mN\left(0 , (1-\pi)\left(\pone T(\nu_{\bts})\right)^2 \sig_{0,\bts,\tht}^2
    \right)
\end{equation}
where
\[
\sig_{0,\bts,\tht}^2 = \intlu (S_{0,\bts}(x))^2 dF(x) - \tht_{0,\bts}^2 = 
\frac{\sig_{\bts,\tht}^2}{\|S_{\bts}\|^2_{G_{\bts}}}.
\]
The asymptotic distribution of the last two summands in \eqref{eq:etahb_expansion} is:
\begin{equation}
\label{eq:eta_hat_b_dist_2}
\begin{aligned}
& 
    \sqrt{\frac{n}{m+n}} 
    \ptwo T(\nut)
    \left[
    \frac{1}{\sqrt{m}}\summ\left(S_{0,\bts}(Y_i) - \del_{0,\bts}\right)
    \right]
    + \\ & \qquad \qquad
    \sqrt{\frac{n}{m+n}} 
    \Gamma_{\bts}^T J_{\bts}^{-1}
    \left[\frac{1}{\sqrt{m}}\summ \pb \log g_{\bts}(Y_i)\right]
    \dconv
    \mN\left(
    0,
    \pi \Lambda_{\bts}
    \right)
\end{aligned}
\end{equation}
where 
\[
    \Lambda_{\bts} = 
    \left(\ptwo T(\nu_{\bts})\right)^2 \sig_{0,\bts,\del}^2 +
    \Gamma_{\bts}^T J_{\bts}^{-1}
    V_{\bts}
    J_{\bts}^{-1} \Gamma_{\bts} + 
    \frac{2\ptwo T(\nu_{\bts})}{\|S_{\bts}\|_{G_{\bts}}}\Gamma_{\bts}^T J_{\bts}^{-1} C_{\bts}
\]
with 
\[
\begin{aligned}
    \sig_{0,\bts,\del}^2 &= 
    \intlu (S_{0,\bt}(x))^2 dF_b(x) - \del_{0,\bts}^2 = \frac{\sig_{\bts,\del}^2}{\|S_{\bts}\|^2_{G_{\bts}}},\\
    V_{\bts} &= Var_{F_b}\Big(\Big[\pb\log g_{\bt(Y)}\Big]_{\bt = \bts}\Big) = \intlu \bigg[\pb\log g_{\bt}(y)\bigg]_{\bt = \bts}\bigg[\pb\log g_{\bt}(y)\bigg]_{\bt = \bts}^TdF_b(y),
\end{aligned}
\]
and
\[
    C_{\bts} = Cov_{F_b}\left(
    \Sd_{\bts}(Y), \Big[\pb\log g_{\bt(Y)}\Big]_{\bt = \bts}\right) = \intlu 
    \Sd_{\bts}(y) \bigg[\pb\log g_{\bt}(y)\bigg]_{\bt = \bts}dF_b(y).
\]
Combining the above expressions, we obtain \eqref{eq:prop_eta_unb_wbkg_beta_est_conv}
and 
\[
\begin{aligned}
    \sig_{\bts,\eta}^2 = & 
    (1-\pi) \pone (T(\nu_{\bts}))^2\sig_{0,\bts,\tht}^2 + 
    \pi\left[
    \ptwo (T(\nu_{\bts}))^2\sig_{0,\bts,\del}^2 + 
    \Gamma_{\bts}^T J_{\bts}^{-1}
    V_{\bts}
    J_{\bts}^{-1} \Gamma_{\bts}  + \frac{2\ptwo T(\nu_{\bts})}{\|S_{\bts}\|_{G_{\bts}}} \Gamma_{\bts}^T J_{\bts}^{-1} C_{\bts}
    \right]
    \\
    &= 
    (1-\pi) A_{\bts}^2 \sig_{\bts,\tht}^2 + \pi \Bigg[
    B_{\bts}^2 \sig_{\bts,\del}^2 + 
    \Gamma_{\bts}^T J_{\bts}^{-1}
    V_{\bts}
    J_{\bts}^{-1} \Gamma_{\bts} + 
    2B_{\bts}\Gamma_{\bts}^T J_{\bts}^{-1} C_{\bts}
    \Bigg]
\end{aligned}
\]
with $A_{\bts} = \frac{1}{(\|S_{\bts}\|_{G_{\bts}}-\del_{\bts})}$ and $B_{\bts} = \frac{(\tht_{\bts} - \|S_{\bts}\|_{G_{\bts}})}{(\|S_{\bts}\|_{G_{\bts}}-\del_{\bts})^2}$.

\hfill $\square$

\subsection{Estimator for variance term in Proposition \ref{prop:eta_unb_wbkg_beta_est}}
\label{app:sig_bts_eta_est}
To derive a consistent estimator for $\sigma_{\bts, \eta}^2$ start by constructing consistent estimators for the quantities $A_{\bts}$, $B_{\bts}$, $J_{\bts}$, $V_{\bts}$, $\Gamma_{\bts}$, and $C_{\bts}$ defined in the previous section.

From Section \ref{app:prop_eta_unb_wbkg_beta_est_proof}, we have $\thth_{0,\bth_m} \pconv \tht_{0,\bts}$ and  $\delh_{0,\bth_m} \pconv \del_{0,\bts}$. By consistency of $\bth_m$ and the dominated convergence theorem, we have $\|S_{\bth_m}\|_{G_{\bth_m}} \pconv \|S_{\bts}\|_{G_{\bts}}$. Hence, $\thth_{\bth_m} \pconv \tht_{\bts}$ and $\delh_{\bth_m} \pconv \del_{\bts}$. Therefore, by the continuous mapping theorem,
\[
    \widehat{A}_{\bth_m} = \frac{1}{\|S_{\bth_m}\|_{G_{\bth_m}} - \delh_{\bth_m}}, 
    \quad\text{and}\quad
    \widehat{B}_{\bth_m} = \frac{\thth_{\bth_m} - \|S_{\bth_m}\|_{G_{\bth_m}}}{(\|S_{\bth_m}\|_{G_{\bth_m}} - \delh_{\bth_m})^2}
\]
are consistent estimators for $A_{\bts}$ and $B_{\bts}$, respectively. 

By consistency of $\bth_m$, $\thth_{\bth_m}$, $\delh_{\bth_m}$ and the law of large numbers, we have
\[
\sigh_{\bth_m, \tht}^2 = \frac{1}{n}\sumn \left(\Sd_{\bth_m}(X_i)\right)^2 - \thth_{\bth_m}^2 \pconv \sig_{\bts, \tht_m}^2 \quad \text{and} \quad
\sigh_{\bth_m, \del}^2 = \frac{1}{m}\summ \left(\Sd_{\bth_m}(Y_i)\right)^2 - \delh_{\bth_m}^2 \pconv \sig_{\bts, \del}^2.
\]

Similar arguments yield the following consistent estimators for $\Gamma_{\bts}, J_{\bts}, V_{\bts}$, and $C_{\bts}$:
\[
\begin{aligned}
    & 
    \widehat\Gamma_{\bth_m} = \left[\frac{1}{(1-\delh_{0,\bth_m})}\left(\frac{1}{n}\sumn\Big[\pb S_{0,\bt}(X_i)\Big]_{\bt = \bth_m}\right) + 
    \frac{(\thth_{0,\bth_m}-1)}{(1-\delh_{0,\bth_m})^2}
    \left(\frac{1}{m}\summ\Big[\pb S_{0,\bt}(Y_i)\Big]_{\bt = \bth_m}\right)
    \right];\\
    &
    \widehat{J}_{\bth_m} = -\frac{1}{m}\summ \Big[\pb^2 \log g_{\bt} (Y_i)\Big]_{\bt = \bth_m};\qquad
    \widehat{V}_{\bth_m} = \frac{1}{m} \summ \bigg[\pb \log g_{\bt}(Y_i)\bigg]_{\bt = \bth_m}\bigg[\pb \log g_{\bt}(Y_i)\bigg]_{\bt = \bth_m}^T;\\
   & \text{and}\quad
    \widehat{C}_{\bth_m} = \frac{1}{m} \summ \Sd_{\bth_m}(Y_i)\bigg[\pb \log g_{\bt}(Y_i)\bigg]_{\bt = \bth_m}.
\end{aligned}
\]
It follows that :
\[
\sigh_{\bth_m, \eta}^2 = 
\frac{m}{m+n} \widehat{A}_{\bth_m}^2\sigh_{\bth_m,\tht}^2 + \frac{n}{m+n}\bigg[
\widehat{B}_{\bth_m}^2\sigh_{\bth_m,\del}^2 + 
\widehat\Gamma_{\bth_m}^T \widehat{J}_{\bth_m}^{-1} \widehat{V}_{\bth_m} \widehat{J}_{\bth_m} \widehat\Gamma_{\bth_m} + 
2 \widehat{B}_{\bth_m} \widehat\Gamma_{\bth_m}^T \widehat{J}_{\bth_m}^{-1}\widehat{C}_{\bth_m}
\bigg] \pconv \sig_{\bts,\eta}^2.
\]
\hfill $\square$

\subsection{Proof of Proposition \ref{prop:thth_0b_dist}}
\label{app:prop_thth_0b_dist_proof}

Recall that $\bt = (\al,\lambda)$ and denote with  $\partial$ and $\partial^2$, respectively,  the gradient vector and the $(p-1) \times (p-1)$ Hessian matrix of a function with respect to $\al$, i.e., the first $(p-1)$ dimensional component of $\beta$.
Under assumptions \ref{assump:gb_cont_bt}-\ref{assump:mB_compact}, the following expansion for $\alh_n$ holds:
\begin{equation}
    \label{eq:al_mle_expansion}
    \sqrt{n}(\alh_n - \als) = \mJ_{\als}^{-1} \frac{1}{\sqrt{n}}\sumn  \big[\p\log q_{{\al}}(X_i)\big]_{\al = \als} + o_P(1)
\end{equation}
where $\mJ_{\als} = \mE_{F}\Big[-\pp\log q_{\al}(X)\Big]_{\al = \als}$. From the definition of $\thth_{0,\bt}$ in \eqref{eq:thth0_b_delh0_b_def} and applying the mean value theorem  about $\bts$ we get:
\begin{equation}
    \label{eq:mvt_al}
    \begin{aligned}
        \thth_{0,\bth_n} =& \thth_{0,\bts} + 
    \Big[\p \thth_{0,\bt}\Big]_{\bt = \tilde\bt}^T (\alh_n - \als);\\
    \sqrt{n}\Big(\thth_{0,\bth_n} - \tht_{0,\bts}\Big) =& 
    \frac{1}{\sqrt{n}}\sumn 
    \Big(S_{0,\bts}(X_i) - \tht_{0,\bts}\Big) + 
    \Big[\p \thth_{0,\bt}\Big]_{\bt = \tilde\bt}^T \sqrt{n}(\alh_n - \als),
    \end{aligned}
\end{equation}
where $\tilde\bt$ is a point lying on the line joining $\bth_n$ and $\bts$. Since $\alh_n$ is a consistent estimator for $\alpha^*$ under $F$, it follows that $\bth_n \to \bts$ and hence $\tilde\bt \to \bts$. Following a similar argument to the one  presented in Section \ref{app:prop_eta_unb_wbkg_beta_est_proof}, \ref{assump:gb_cont_x}-\ref{assump:bts_interior} yield:
\begin{equation}
    \label{eq:pal_thth_conv}
    \Big[\p \thth_{0,\bt}\Big]_{\bt = \btt} = \frac{1}{n} \sumn \Big[\p S_{0,\bt}(X_i)\Big]_{\bt = \btt} \pconv
    \intlu \Big[\p S_{0,\bt}(x)\Big]_{\bt = \bts}dF(x) = 
    \Big[\p \tht_{0,\bt}\Big]_{\bt = \bts}.
\end{equation}
Combining \eqref{eq:al_mle_expansion}-\eqref{eq:pal_thth_conv} leads to:
\begin{equation}
    \label{eq:thth_dist_cov}
    \begin{aligned}
        \sqrt{n}\Big(\thth_{0,\bth_n} - \tht_{0,\bts}\Big)  =& 
    \frac{1}{\sqrt{n}}\sumn 
    \Big(S_{0,\bts}(X_i) - \tht_{0,\bts}\Big) + \\
    &
    \Big[\p \tht_{0,\bt}\Big]_{\bt = \bts}^T
    \mJ_{\als}^{-1} \frac{1}{\sqrt{n}}\sumn \bigg[\p \log q_{\al}(X_i)\bigg]_{\al = \als} + o_P(1)
    \end{aligned}
\end{equation}
and, by applying the central limit theorem on \eqref{eq:thth_dist_cov}, we obtain \eqref{eq:prop_thth_0b_dist} with
\[
\begin{aligned}
    \sig_{\bts,\tht_0}^2 =& 
\Big[\intlu \big(S_{0,\bts}(x)\big)^2dF(x) - \tht_{0,\bts}^2\Big] + D_{\als}^T\mJ_{\als}^{-1}\mV_{\als}\mJ_{\als}^{-1}D_{\als} + 
2D_{\als}^T\mJ_{\als}^{-1}\mC_{\als}\\
=&
\frac{\sig^2_{\bts,\tht}}{\|S_{\bts}\|^2_{G_{\bts}}} + D_{\als}^T\mJ_{\als}^{-1}\mV_{\als}\mJ_{\als}^{-1}D_{\als} + 
2D_{\als}^T\mJ_{\als}^{-1}\mC_{\als}
\end{aligned}
\]
where 
\[
\begin{aligned}
    \mV_{\als} =& Var_{F}\Big(\bigg[\p \log q_{\al}(X)\bigg]_{\al = \als}\Big) = \intlu \Big[\p\log q_{\al}(x)\Big]_{\al = \als}\Big[\p\log q_{\al}(x)\Big]_{\al = \als}^TdF(x),\\
    \mC_{\als} =& Cov_F\Big(S_{0,\bts}(X), \bigg[\p \log q_{\al}(X)\bigg]_{\al = \als}\Big) = 
    \intlu S_{0,\bts}(x)\Big[\p \log q_{\al}(x)\Big]_{\al = \als}dF(x)
\end{aligned}
\]
and $D_{\als} = \Big[\p \tht_{0,\bt}\Big]_{\bt = \bts}$.
\hfill $\square$

\subsection{Estimator for variance term in Proposition \ref{prop:thth_0b_dist}}
\label{app:sig_bts_tht0_est}
We start by constructing consistent estimators for $D_{\als}, \mJ_{\als}, \mV_{\als}$ and $\mC_{\als}$ as defined in the previous section. From \eqref{eq:mvt_al}, we have $\thth_{0,\bth_n} \pconv \tht_{0,\bts}$. By consistency of $\alh_n$ and the dominated convergence theorem, we get:
\[
\|S_{\bth_n}\|_{G_{\bth_n}} \pconv \|S_{\bts}\|_{G_{\bts}} \quad \text{and} \quad
\sigh_{\bth_n,\tht}^2 = \frac{1}{n} \sumn \Big(\Sd_{\bth_n}(X_i)\Big)^2 - \thth_{\bth_n}^2 \pconv \sig_{\bts,\tht}^2.
\]
Similarly, we have the following consistent estimators for $D_{\als}, \mJ_{\als}, \mV_{\als}$ and $\mC_{\als}$:
\[
\begin{aligned}
    \widehat{D}_{\alh_n} =& \frac{1}{n} \bigg[\p S_{0, \bt}(X_i)\bigg]_{\bt = \bth_n};\qquad
    \widehat{J}_{\alh_n} = -\frac{1}{n} \sumn \bigg[\pp \log q_{\al}(X_i)\bigg]_{\al = \alh_n};\\
    \widehat{\mV}_{\alh_n} =& \frac{1}{n} \sumn \Big[\p \log q_{\al}(X_i)\Big]_{\al = \alh_n}\Big[\p \log q_{\al}(X_i)\Big]_{\al = \alh_n}^T;\\
    \text{and}\quad& \widehat{\mC}_{\alh_n} = \frac{1}{n} \sumn 
    S_{0, \bth_n}(X_i)
    \Big[\p \log q_{\al}(X_i)\Big]_{\al = \alh_n}.
\end{aligned}
\]
Thus, by the continuous mapping theorem, it follows that,
\[
\sigh_{\bth_n, \tht_0}^2 = \frac{\sigh_{\bth_n,\tht}^2}{\|S_{\bth_n}\|_{G_{\bth_n}}^2} + 
\widehat{D}_{\alh_n}^T\widehat{\mJ}_{\alh_n}^{-1}\widehat{\mV}_{\alh_n}\widehat{\mJ}_{\alh_n}^{-1}\widehat{D}_{\alh_n} + 2\widehat{D}_{\alh_n}^T \widehat{\mJ}_{\alh_n}^{-1}\widehat{\mC}_{\alh_n} \pconv 
\sig_{\bts, \tht_0}^2.
\]
\hfill $\square$

\newpage

\section{Additional figures}
\label{app:additional figures}

\begin{figure}[!h]
    \includegraphics[width=0.45\textwidth]{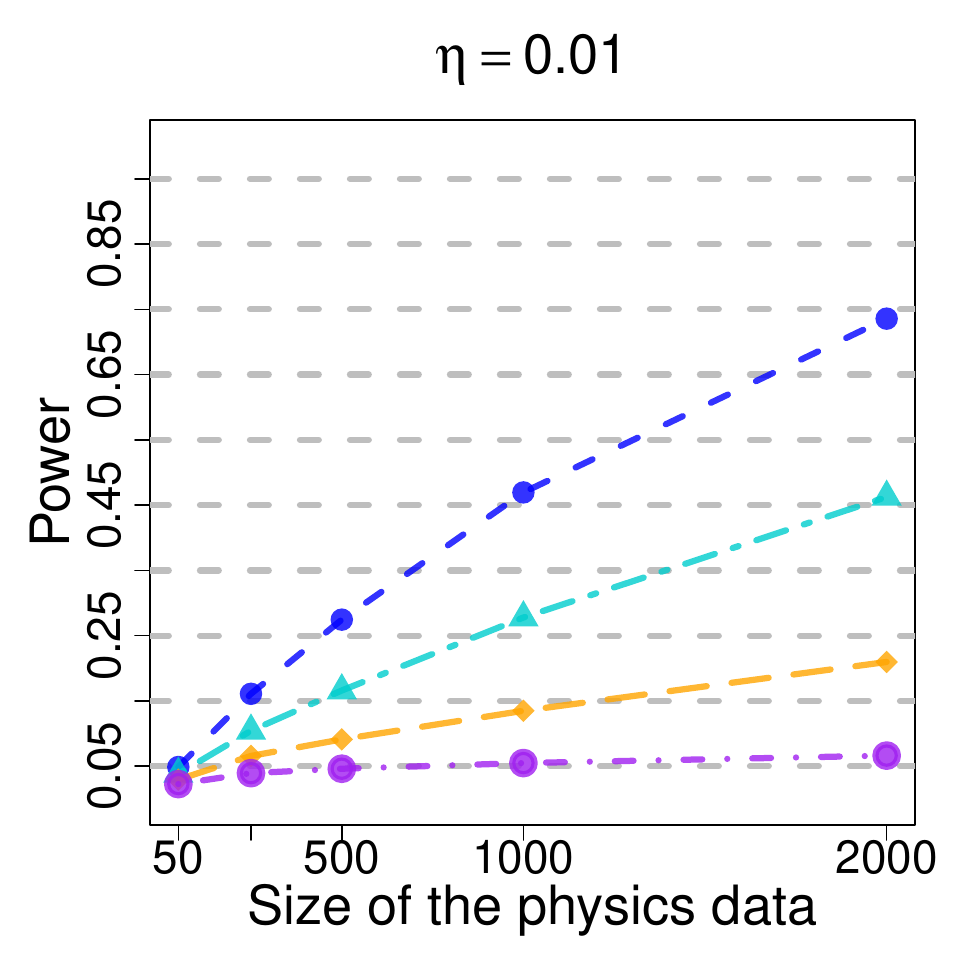}
       \includegraphics[width=0.45\textwidth]{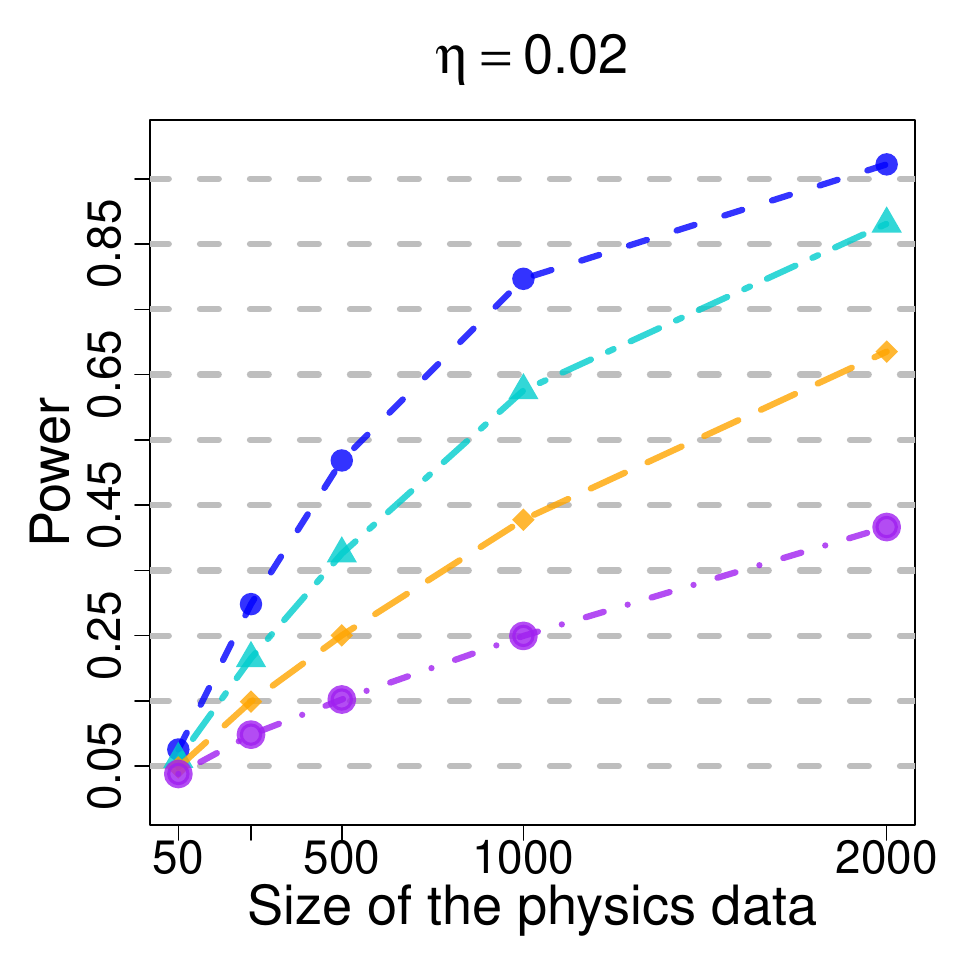}
\centering  \begin{subfigure}[b]{\textwidth}
    \centering
    \includegraphics[width=0.7\textwidth, height = 2cm]{power_curve_legend_wobkg.pdf}
  \end{subfigure}
  \caption{Simulated power curves for the numerical example in Section \ref{sec:inf_wobkg} with $\eta = 0.01$ (left panel), $\eta = 0.02$ (right panel), and letting $\lambda$ and $n$ vary.} 
  \label{fig:power_curve_wobkg_2}
\end{figure}

\bibliographystyle{apalike}
\bibliography{final_ref}
\end{document}